\documentclass[final,3p,12pt,times]{elsarticle}
\usepackage{blkarray}
\usepackage{amssymb}
\usepackage{amsmath}
\usepackage{booktabs}
\usepackage{multirow}
\usepackage{xcolor}
\usepackage{subfigure}
\usepackage{array}
\usepackage{setspace}
\usepackage{amsthm}
\usepackage{float}
\usepackage{caption}
\usepackage[figuresright]{rotating}
\usepackage{longtable}
\usepackage{tabularx}
\usepackage{makecell}
\usepackage{algorithm}
\usepackage{algpseudocode}
\usepackage{tikz}
\usepackage{graphicx}
\usepackage{changepage}
\usepackage{etoolbox}
\usepackage{adjustbox}
\usepackage{subcaption}
\AtBeginEnvironment{proof}{\begin{adjustwidth}{1.5em}{}}
\AtEndEnvironment{proof}{\end{adjustwidth}}
\makeatletter

\newcommand{\Rmnum}[1]{\expandafter\@slowromancap\romannumeral #1@}
\makeatother

\definecolor{MatrixPurple}{RGB}{138,43,226} 
\definecolor{MatrixBlue}{RGB}{34,139,34} 
\definecolor{MatrixRed}{RGB}{176,23,31} 
\journal{Signal Processing}

\begin{document}
\begin{sloppypar}
	
\begin{frontmatter}



\title{Trainable Joint Time-Vertex Fractional Fourier Transform}

\author[1]{Ziqi~Yan}
 \author[1,2,3,4]{Zhichao Zhang\corref{cor1}}

\ead{zzc910731@163.com}
\cortext[cor1]{Corresponding author; Tel: +86-13376073017.}
\address[1]{School of Mathematics and Statistics, Nanjing University of Information Science and Technology, Nanjing 210044, China}
\address[2]{Hubei Key Laboratory of Applied Mathematics, Hubei University, Wuhan 430062, China}
\address[3]{Key Laboratory of System Control and Information Processing, Ministry of Education, Shanghai Jiao Tong University, Shanghai 200240, China}
\address[4]{Key Laboratory of Computational Science and Application of Hainan Province, Hainan Normal University, Haikou 571158, China}

\tnotetext[mytitlenote]{This work was supported in part by the Open Foundation of Hubei Key Laboratory of Applied Mathematics (Hubei University) under Grant HBAM202404; in part by the Foundation of Key Laboratory of System Control and Information Processing, Ministry of Education under Grant Scip20240121; and in part by the Foundation of Key Laboratory of Computational Science and Application of Hainan Province under Grant JSKX202401.}


%
\begin{abstract}
To address limitations of the graph fractional Fourier transform (GFRFT) Wiener filtering and the traditional joint time-vertex fractional Fourier transform (JFRFT)
Wiener filtering, this study proposes a filtering method based on the hyper-differential form of the JFRFT. The gradient backpropagation mechanism is employed to enable the adaptive selection of transform order pair and filter coefficients. First, leveraging the hyper-differential form of the GFRFT and the fractional Fourier transform, the hyper-differential form of the JFRFT is constructed and its properties are analyzed. Second, time-varying graph signals are divided into dynamic graph sequences of equal span along the temporal dimension. A spatiotemporal joint representation is then established through vectorized reorganization, followed by the joint time-vertex Wiener filtering. Furthermore, by rigorously proving the differentiability of the transform orders, both the transform orders and filter coefficients are embedded as learnable parameters within a neural network architecture. Through gradient backpropagation, their synchronized iterative optimization is achieved, constructing a parameters-adaptive learning filtering framework. This method leverages a model-driven approach to learn the optimal transform order pair and filter coefficients. Experimental results indicate that the proposed framework improves the time-varying graph signals denoising performance, while reducing the computational burden of the traditional grid search strategy.
\end{abstract}



\begin{keyword}
Graph signal denoising \sep Graph fractional Fourier transform \sep Joint time-vertex  \sep Machine learning



\end{keyword}

\end{frontmatter}


\section{Introduction}
Graph signal denoising is a fundamental task in graph signal processing (GSP) and is extensively used in fields such as medical imaging, financial data analysis, bioinformatics and communication networks. In 
 the traditional signal denoising, the Fourier transform (FT) is employed to map signals from the time domain to the frequency domain, where noise is effectively removed using designed filters. This approach has been extended to graph signal denoising, where the graph Fourier transform (GFT) is used to convert graph signals from the vertex domain to the graph spectral domain \cite{ref1,ref2,ref3,ref4}, enabling noise removal through spectral domain filtering.

Recent advancements in GFT have introduced the concept of transform order, leading to the development of the graph fractional Fourier transform (GFRFT) \cite{ref5,ref6,ref7,ref8,ref9}. The GFRFT enables the transformation of graph signals from the vertex domain to the intermediate vertex-spectral domains, facilitating more effective denoising through filtering in these intermediate domains. Wang et al. initially proposed the fractional power definition of the GFRFT, defining the transform order within a finite real interval \cite{ref10}. Extending this theory, Alikaşifoğlu et al. introduced a hyper-differential operator definition for the GFRFT and expanded the transform order to the entire real domain \cite{ref11}.

In practical scenarios, graph signals often exhibit time-varying characteristics, meaning that data associated with graph vertices change over time. This dynamic nature limits the effectiveness of the traditional static graph signal processing methods, which do not adequately capture temporal features. To address this, Loukas et al. integrated discrete signal processing with GSP by combining the GFT and the FT, leading to the development of the joint time-vertex Fourier transform (JFT) \cite{ref12,ref13}. This method facilitated synchronized time-vertex domain analysis of time-varying graph signals and contributed to a range of applications, including time-varying graph signal reconstruction \cite{ref14,ref15,ref16} and semi-supervised learning \cite{ref17,ref18}. Building on this foundation, Alikaşifoğlu et al. combined the fractional power definition of the GFRFT with the fractional Fourier transform (FRFT), introducing the joint time-vertex fractional Fourier transform (JFRFT) \cite{ref19}, which provided a more flexible framework for denoising time-varying graph signals.

In the classical signal denoising, Wiener filtering, based on the minimum mean square error criterion, has proven to be highly effective. This success motivated researchers to extend Wiener filtering to graph signal denoising \cite{ref20}. Gavili et al. were the first to apply the frequency domain Wiener filtering in the graph spectral domain, denoising static graph signals through spectral domain filtering \cite{ref21}. Ozturk et al. further extended the spectral domain Wiener filtering to the GFRFT domain, selecting different transform orders, with prior knowledge of the graph signal and noise used to compute optimal filter coefficients, thereby enhancing denoising performance compared to the traditional graph spectral Wiener filtering \cite{ref22}. Subsequently, Alikaşifoğlu et al. employed the hyper-differential form of the GFRFT and proposed two methods for denoising time-varying graph signals \cite{ref11}. In the first method,  selecting different transform orders, with the mean of auto-correlation and cross-correlation over time used to generate the optimal filter. The second method employs a gradient-based adaptive selection mechanism that leverages the differentiability of the GFRFT. By utilizing gradient backpropagation, this method adaptively learns the transform order and filter coefficients, further optimizing the denoising process.

Despite the effectiveness of GFRFT-based filtering, it does not fully account for the temporal characteristics of time-varying graph signals. To address this limitation, Alikaşifoğlu et al. investigated Wiener filtering in the JFRFT domain \cite{ref23}. They derived filter coefficients for different transform order pairs and employed a grid search method to determine the most suitable pair. The corresponding filter coefficients were then computed, facilitating effective denoising of time-varying graph signals in the JFRFT domain.

Existing Wiener filtering methods for time-varying graph signals present two key limitations. First, in GFRFT-based filtering, while the gradient backpropagation mechanism enables adaptive learning of transform order and filter coefficients, thereby enhancing denoising efficiency, the temporal characteristics of time-varying graph signals remain insufficiently addressed, leaving room for performance improvement. Second, in JFRFT-based filtering, although temporal characteristics are incorporated, the reliance on a grid search strategy results in excessive computational costs, necessitating a more efficient approach. To overcome these limitations, this study builds on the adaptive learning framework of transform order and filter coefficients in GFRFT-based filtering while integrating temporal characteristics. A learnable JFRFT is constructed based on the hyper-differential form of the GFRFT and the FRFT. By embedding the transform order pair and filter coefficients into a neural network framework and utilizing gradient backpropagation, the proposed method adaptively learns optimal transform order pair and filter coefficients, enabling more effective denoising of time-varying graph signals.

The primary contributions of this study are as follows:
\begin{itemize}
 \item The existing JFRFT framework is extended to the learnable JFRFT based on hyper-differential operators, with a comprehensive analysis of its properties.
 \item A novel method is introduced to optimize transform order pair by learning their values directly from task-specific data.
 \item Through a gradient backpropagation mechanism, transform order pair and filter coefficients are adaptively optimized, significantly reducing computational costs while enhancing denoising efficiency for time-varying graph signals.
 \item Extensive experiments on synthetic and real-world datasets validate the effectiveness of the proposed denoising method, demonstrating its superiority over existing approaches.
\end{itemize}

The distinctions and connections between this study and existing research are summarized as follows:
\begin{itemize}
\item Building on the gradient-based adaptive learning framework in GFRFT filtering methods, this study develops the hyper-differential form of JFRFT to enable adaptive filtering for time-varying graph signals. Unlike the GFRFT approach, the proposed method further incorporates temporal features, thereby enhancing denoising performance.
\item The fractional power form of JFRFT is extended to a hyper-differential form, generalizing the transform order pairs to the entire real domain. Existing JFRFT methods rely on grid search for transform order pairs selection, resulting in high computational costs. To address this limitation, the proposed approach introduces a gradient-based adaptive learning mechanism, eliminating the need for grid search and improving  algorithmic efficiency.
\end{itemize}

The structure of this paper is as follows: Section 2 presents the theoretical foundations, Section 3 demonstrates the learnability of the model  and Section 4 validates the effectiveness of the proposed method through numerical experiments, comparing it with existing time-varying graph filtering techniques.

Notation: In the following text, bold lowercase letters represent vectors and bold uppercase letters denote matrices. The symbol \(\lvert \mathcal{V} \rvert\) denotes the cardinality of the set \(\mathcal{V}\). The operations of complex conjugate, transpose and the Hermitian transpose are denoted by \((\cdot)^*\), \((\cdot)^T\) and \((\cdot)^H\). Additionally, \(\lVert \textbf{A} \rVert_F\) denotes the Frobenius norm of the matrix \textbf{A}, \( \| \cdot \|_0 \) represents the \( \ell_0 \) pseudo-norm, $\textbf{I}_N$ denotes the identity matrix of size $N$. The operator vec(·) denotes vectorization,
obtained by stacking the matrix columns. For $\textbf{A} \in \mathbb{C}^{m \times n}$ and $\textbf{B} \in \mathbb{C}^{p \times q}$, $\textbf{A} \otimes \textbf{B} \in \mathbb{C}^{mp \times nq}$ denotes their Kronecker product. For the same size $\mathbf{A}$ and $\mathbf{B}$, $\mathbf{A} \odot \mathbf{B}$ denotes their Hadamard product.

\section{Preliminaries} \label{Sec:Preliminaries}
In this section, we first review the fundamental concepts of graph signals and the GFT, then discuss the GFRFT and the FRFT.

\subsection{Graph Signals and GFT}
Considering a graph \(\mathcal{G} = (\mathcal{V}, \mathbf{A})\), where \(\mathcal{V} = \{ v_1, v_2, \dots, v_n \}\) and \(|\mathcal{V}| = N\) represents the set of vertices. The matrix \(\mathbf{A} \in \mathbb{C}^{N \times N}\) is the weighted adjacency matrix of the graph. If there is an edge connecting node \(n\) to node \(m\), then \(\mathbf{A}_{m,n} \neq 0\), where \(\mathbf{A}_{m,n}\) denotes the element in the \(m\)-th row and \(n\)-th column of the matrix \(\mathbf{A}\). In contrast, if \(\mathbf{A}_{m,n} = 0\), it denotes that there is no edge connecting node \(n\) and node \(m\). A graph is considered undirected if \(\mathbf{A}_{m,n} = \mathbf{A}_{n,m}\) for all \(m, n \in \{1, \dots, N\}\) \cite{ref24}.

Let \(\mathbf{x} = [x_1, x_2, \dots, x_N]^T\) represent a graph signal, where \(\mathbf{x} \in \mathbb{C}^N\) is a mapping from the set of vertices \(\mathcal{V}\) to the complex number field \(\mathbb{C}\). Each vertex \(v_n\) corresponds to a complex number \(x_n\), such that \(v_n \mapsto x_n\) for all \(n \in \{1, \dots, N\}\).

The GFT is defined through the Jordan decomposition of the graph shift operator matrix \( \mathbf{Z} \in \mathbb{C}^{N \times N} \). This decomposition provides a general framework for various graph shift operators, such as the adjacency matrix \( \mathbf{A} \), the Laplacian matrix \( \mathbf{L} \), the row-normalized adjacency matrix \( \mathbf{Q} = \mathbf{D}^{-1} \mathbf{A} \) and the symmetric normalized Laplacian \( \boldsymbol{\mathcal{L}} = \mathbf{D}^{-1/2} \mathbf{L} \mathbf{D}^{-1/2} \). Let \( \mathbf{x} \) be a graph signal on a graph \( \mathcal{G} \) with the graph shift operator matrix \( \mathbf{Z} \). We can express \( \mathbf{Z} = \mathbf{V} \mathbf{J}_Z \mathbf{V}^{-1} \), where \( \mathbf{J}_Z \) is the Jordan normal form of \( \mathbf{Z} \) and \( \mathbf{V} = [\textbf{v}_1, \textbf{v}_2, \cdots, \textbf{v}_N] \) contains the generalized eigenvectors of \( \mathbf{Z} \) as its columns. The GFT matrix is defined as \( \textbf{F}_G = \textbf{V}^{-1} \), which allows us to compute the GFT of \( \textbf{x} \) as follows \cite{ref11}:
\begin{equation}
\tilde{\textbf{x}} \triangleq \textbf{F}_G \textbf{x} = \textbf{V}^{-1} \textbf{x}.
\end{equation}
where \( \tilde{\textbf{x}} \) represents the graph signal in the GFT domain. The inverse transformation to return to the vertex domain is given by:
\begin{equation}
\textbf{x} = \textbf{F}_G^{-1} \tilde{\textbf{x}} = \textbf{V} \tilde{\textbf{x}}.
\end{equation}

\subsection{GFRFT}

The fractional power-based definition of the GFRFT generalized the FRFT to the GSP domain \cite{ref25, ref26,ref27}. In \cite{ref11}, the author extended this definition to support any graph structure. Let \( \textbf{Z} \) represent a graph shift operator matrix of an arbitrary graph \( \mathcal{G} \), which can be expressed in its Jordan decomposition as \( \textbf{Z} = \textbf{V}_Z \textbf{J}_Z \textbf{V}_Z^{-1} \). Here, the GFT matrix is defined as $\textbf{F}_G={\textbf{V}}_Z^{-1}=\textbf{P}\textbf{J}_F{\textbf{P}}^{-1}$, \( \textbf{J}_F \) is the Jordan form of ${\textbf{F}}_G$ and $\textbf{P}$ is the corresponding matrix whose columns contain the generalized eigenvectors of ${\textbf{F}}_G$. Thus, the GFRFT with order \( \alpha \) is defined as:
\begin{equation}
\textbf{F}_G^{\alpha} = \textbf{P} \textbf{J}_F^{\alpha} \textbf{P}^{-1}, \quad \alpha \in [0, 1].
\end{equation}

Subsequently, the author presented a hyper-differential operator definition for the GFRFT. This derivation is consistent with hyper-differential operator-based definition of FRFT and supports any transform orders.  For any \( \alpha \in \mathbb{R} \), the GFRFT is defined as:
\begin{equation}
    \begin{gathered}
        \mathbf{F}_{G}^{\alpha} = \exp\left(-j\frac{\alpha\pi}{2} 
        \left(\pi\left(\mathbf{D}_{G}^{2} + \mathbf{F}_{G}\mathbf{D}_{G}^{2}\mathbf{F}_{G}^{-1}\right) - \frac{1}{2}\mathbf{I}\right)\right) ,\\
        \mathbf{D}_{G}^{2} = \frac{1}{2\pi}\left(\frac{j2}{\pi}\log(\mathbf{F}_{G}) + \frac{1}{2}\mathbf{I}\right).
    \end{gathered}
\end{equation}

The two definitions of the GFRFT are equivalent, exhibit index additivity and reduce to \( \mathbf{I} \) for \( \alpha = 0 \) and ordinary \( \mathbf{F}_G \) for \( \alpha = 1 \).
\subsection{FRFT}
The FRFT is a generalization of the FT with a free parameter \( \alpha \). The $\alpha$-th order FRFT of a continuous time signal $x(t)$ is defined as follows:

\begin{equation}
    \begin{aligned}
        X_{\alpha}(u) &= \mathcal{F}^{\alpha}[x](u) \triangleq \int_{-\infty}^{+\infty} x(t) K_{\alpha}(t, u) \, dt, \\
        K_{\alpha}(t, u) &\triangleq \left\{
        \begin{array}{ll}
            A_{\alpha} e^{j \cot \frac{\alpha}{2} t^2 - j u t \csc \frac{\alpha}{2} + j \cot \frac{\alpha}{2} u^2},\alpha \neq n\pi \\
            \delta(t - u),\alpha = 2n\pi  \\
            \delta(t + u),\alpha = (2n \pm 1)\pi 
        \end{array}
        \right.,\\
    \end{aligned}
\end{equation}
where $A_{\alpha} \triangleq \sqrt{\frac{1 - j \cot \alpha}{2\pi}}$ and $n \in \mathbb{Z}$, $\delta (\cdot)$ being the Dirac delta function \cite{ref28}.

For the discrete fractional Fourier transform (DFRFT), using eigen decomposition method, the DFRFT matrix is defined as follows:
\begin{equation}
\textbf{F}^{\alpha}_{m,n} = \sum_{\substack{k=0 \\ k \neq N-1 + (N)_2}}^{N} \boldsymbol{\varphi}^{(k)}_m e^{-j \frac{\alpha \pi}{2} k} \boldsymbol{\varphi}^{(k)}_n,
\end{equation}
 where \( (N)_2 \equiv N \mod 2 \), \( \boldsymbol{\varphi}^{(k)}_n \) is the entry \( n \) of the \( k \)$-$th discrete Hermite-Gaussian vector \cite{ref29}.
 
Hyper-differential operator theory can also be used to build the discrete DFRFT matrix. The matrix is defined as follows:

\begin{equation}
\textbf{F}^\alpha = \exp \left[ -j \frac{\pi \alpha}{2} \left( \pi (\textbf{U}^2 + \textbf{D}^2) - \frac{1}{2} \textbf{I} \right) \right],
\end{equation}
where \( \textbf{D} \) and \( \textbf{U} \) are the discrete manifestations of the differentiation and coordinate multiplication operators, respectively. Here, the definitions in (6) and (7) are equivalent \cite{ref30,ref31}, exhibit index additivity and reduce to \( \mathbf{I} \) for \( \alpha = 0 \) and FT for \( \alpha = 1 \).

\section{Learnable JFRFT}
In our study, for transform order pair $(\alpha, \beta) \in \mathbb{R}^2$, we use hyper-differential operator-based GFRFT and DFRFT to define learnable JFRFT operator as follows:
\begin{equation}
    \mathbf{F}_{J}^{\alpha,\beta} \triangleq \mathbf{F}^{\beta} \otimes \mathbf{F}_{G}^{\alpha}.
\end{equation}
\subsection{Properties}

In what follows, we present some properties of the proposed learnable JFRFT based on a hyper-differential operator. Among them, properties (4), (5) and (6) can be readily demonstrated based on the established properties of (3).

(1) Reduction to the identity for $(\alpha, \beta) = (0,0)$:
\begin{equation}
\mathbf{F}_{J}^{0,0} = \textbf{F}^{0} \otimes \textbf{F}_{G}^{0} = \textbf{I}_{T} \otimes \textbf{I}_{N} = \textbf{I}_{NT}.
\end{equation}

(2) Reduction to the ordinary transform for $(\alpha, \beta) = (1,1)$:
\begin{equation}
\mathbf{F}_{J}^{1,1} = \textbf{F}^{1} \otimes \textbf{F}_{G}^{1} = \textbf{F}_{J}.
\end{equation}

(3) Index additivity:
\begin{equation}
\mathbf{F}_{J}^{\alpha_1, \beta_1} \mathbf{F}_{J}^{\alpha_2, \beta_2} = \mathbf{F}_{J}^{\alpha_1 + \alpha_2, \beta_1 + \beta_2}.
\end{equation}

\begin{proof}
Considering a time-varying graph signal \(\textbf{X} \in \mathbb{C}^{N \times T}\):
\begin{equation}
\begin{aligned}
\mathbf{F}_{J}^{\alpha_1, \beta_1} \mathbf{F}_{J}^{\alpha_2, \beta_2} \mathbf{X} 
&= \mathbf{F}_{G}^{\alpha_1} \mathbf{F}_{G}^{\alpha_2} \mathbf{X} \left( \mathbf{F}^{\beta_2} \right)^T \left( \mathbf{F}^{\beta_1} \right)^T \\
&= \mathbf{F}_{G}^{\alpha_1} \mathbf{F}_{G}^{\alpha_2} \mathbf{X} \left( \mathbf{F}^{\beta_1} \mathbf{F}^{\beta_2} \right)^T\\
&= \mathbf{F}_{G}^{\alpha_1 + \alpha_2}\textbf{X}\left({\textbf{F}}^{\beta_1 + \beta_2}\right)^T.
\end{aligned}
\end{equation}
\end{proof}

(4) Commutative:
\begin{equation}
\mathbf{F}_{J}^{\alpha_1, \beta_1} \mathbf{F}_{J}^{\alpha_2, \beta_2} = \mathbf{F}_{J}^{\alpha_1, \beta_2} \mathbf{F}_{J}^{\alpha_2, \beta_1} = \mathbf{F}_{J}^{\alpha_2, \beta_1} \mathbf{F}_{J}^{\alpha_1, \beta_2}.
\end{equation}

(5) Reversible:
\begin{equation}
 \mathbf{F}_{J}^{-\alpha, -\beta} \mathbf{F}_{J}^{\alpha, \beta} = \mathbf{F}_{J}^{-\alpha+\alpha, -\beta+\beta} = \mathbf{F}_{J}^{0, 0} = \mathbf{I_{NT}}.
\end{equation}

(6) Separable:
\begin{equation}
\mathbf{F}_{J}^{\alpha, \beta} = \mathbf{F}_{J}^{\alpha+0, 0+\beta} = \mathbf{F}_{J}^{\alpha, 0} \mathbf{F}_{J}^{0, \beta} =  \mathbf{F}_{G}^{\alpha} \mathbf{F}^{\beta}.
\end{equation}

\subsection{Bandlimitedness}
Bandlimitedness has long been a fundamental concept in the traditional signal processing. The study of bandlimited signals in the FRFT domain can be found in \cite{ref32,ref33}. In recent years, the concept of bandlimitedness has been expanded to GSP \cite{ref34,ref35}.

\textbf{Bandlimitedness of GFT:} According to the definition provided in \cite{ref36,ref37}, a graph signal \( \textbf{x} \) with \( N \) nodes is considered bandlimited if its GFT, denoted as \( \tilde{\textbf{x}} = \textbf{F}_G \textbf{x} \), is \( K \)-sparse, where \( K \leq N \) and \( \| \tilde{\textbf{x}} \|_0 \leq K \). Furthermore, it is shown in \cite{ref37}, without loss of generality, we can assume that the last \( N - K \) entries in \( \tilde{\textbf{x}} \) are zero, i.e., \( \tilde{\textbf{x}} = [\tilde{x}_K^T, \mathbf{0}_{N-K}^T]^T \).

\textbf{Bandlimitedness of GFRFT:} Extending the bandlimitedness to the GFRFT, as proposed in \cite{ref25}, a graph signal \( \textbf{x} \) with \( N \) nodes is considered \( \alpha \)-bandlimited if its GFRFT, denoted as \( \tilde{\textbf{x}}_{\alpha} = \textbf{F}_G^{\alpha} \textbf{x} \), is \( K \)-sparse, where \( K \leq N \) and \( \|  \tilde{\textbf{x}}_\alpha  \|_0 \leq K \). Without loss of generality, we can also assume that the last \( N - K \) entries in \( \tilde{\textbf{x}}_\alpha \) are zero.

\textbf{Bandlimitedness of Learnable JFRFT:} The previous explanations focus on the bandlimitedness of static graph signals. We now extend our analysis to the bandlimitedness of time-varying graph signals. Let \(\textbf{X}\) represents a \( K \)-\( L \) bandlimited graph signal with \( N \) vertices and \( T \) time instances, where \( K \leq N \) and \( L \leq T \). Thus, we can express \(\tilde{\textbf{X}} = {\textbf{F}}_G^{\alpha} \textbf{X} \left( \mathbf{F}^{\beta} \right)^T\), from the row perspective, \(\tilde{\textbf{X}}\) is a matrix where first \( K \) rows are \( L \)-sparse vectors and from the column perspective, first \( L \) columns are \( K \)-sparse vectors. Below, to present clearly, we present a time-varying graph signal that is \( K \)-\( L \) bandlimited in the (\(\alpha, \beta\)) domain.

\[
\underset{\text{\scriptsize Bandlimited graph signal in \(\left(\alpha,\beta\right)\) domain}}{\left( 
\begin{array}{ccccc}
a_{1,1} & a_{1,2} & \cdots & a_{1,L} & 0 \cdots 0 \\
a_{2,1} & a_{2,2} & \cdots & a_{2,L} & 0 \cdots 0 \\
\vdots  & \vdots  & \ddots & \vdots  & \vdots \\
a_{K,1} & a_{K,2} & \cdots & a_{K,L} & 0 \cdots 0 \\
0 & 0 & \cdots & 0 & 0 \cdots 0 \\
\vdots & \vdots & \ddots& \vdots & \vdots \\
0 & 0 & \cdots & 0 & 0 \cdots 0 \\
\end{array}
\right)_{\!\!N \times T}}
\]

\subsection{ Learnable Transform Orders Approach}

The DFRFT matrix is differentiable with respect to the fractional order. Building on this differentiability in \cite{ref38}, the author demonstrates that the fractional order can be integrated into neural network architectures as a trainable layer. During the training stage, the transform order is learned using backpropagation. This success motivated researchers to extend the idea to the GFRFT \cite{ref11}, allowing the fractional order of the GFRFT to be learned similarly. We further extend the approach to the JFRFT, where the transform order pair, along with other network weights, can also be learned through backpropagation within the neural network framework. In equations (16) and (17) presented below, we show the differentiability of DFRFT and GFRFT.

\begin{equation}
\begin{aligned}
\dot{\textbf{F}}^{\beta} &= \frac{d}{d\beta} \textbf{F}^{\beta} = \frac{d}{d\beta} \exp \left( -j \frac{\pi\beta}{2} \left( \pi(\textbf{U}^2 + \textbf{D}^2) - \frac{1}{2}\textbf{I} \right) \right) \\
&= \frac{d}{d\beta} \exp \left( \beta \textbf{T} \right) = \textbf{T} \exp(\beta \textbf{T}), \quad \text{where} \ \textbf{T} = -j \frac{\pi}{2} \left( \pi(\textbf{U}^2 + \textbf{D}^2) - \frac{1}{2}\textbf{I} \right).
\end{aligned}
\end{equation}

\begin{equation}
\begin{aligned}
\dot{\textbf{F}}_G^{\alpha} &= \frac{d}{d\alpha} {\textbf{F}}_{G}^{\alpha} = \frac{d}{d\alpha} \exp\left(-j\frac{\pi\alpha}{2} 
        \left(\pi\left(\mathbf{D}_{G}^{2} + \mathbf{F}_{G}\mathbf{D}_{G}^{2}\mathbf{F}_{G}^{-1}\right) - \frac{1}{2}\mathbf{I}\right)\right) \\
&= \frac{d}{d\alpha} \exp \left( \alpha \tilde{\textbf{T}} \right) = \tilde{\textbf{T}} \exp(\alpha \tilde{\textbf{T}}), \quad \text{where} \ \tilde{\textbf{T}} = -j \frac{\pi}{2} \left(\pi\left(\mathbf{D}_{G}^{2} + \mathbf{F}_{G}\mathbf{D}_{G}^{2} \mathbf{F}_{G}^{-1} -  \frac{1}{2} \mathbf{I} \right)\right).
\end{aligned}
\end{equation}

In deep learning architectures, the JFRFT can be regarded as a fully connected layer that preserves the same input and output dimensions. Unlike conventional fully connected layers, all weights are determined by two parameters: the graph fractional order \( \alpha \) and the fractional order \( \beta \). The relationship between the activations of the \( \ell \)-th layer and the \( (\ell - 1) \)-th layer is expressed as $\textbf{x}^{(\ell)} = \varphi \left( {\textbf{F}}_{J}^{\alpha,\beta} \textbf{x}^{(\ell - 1)} \right)$. Where \( \varphi \) denotes an arbitrary differentiable activation function. In this framework, both the graph fractional order and the fractional order can be treated as two learnable parameters within the neural network, which can be trained to optimize the transform order pair for a specific task.

Let \( \mathcal{L} \) denote the defined loss function for the network, which depends on the activations of the \( \ell \)-th layer, represented as \( \mathcal{L}(\textbf{x}^{(\ell)}) \). The gradient of the loss function with respect to the activations of the \( \ell \)-th layer is denoted by $\nabla_\mathbf{x}^{(\ell)} \mathcal{L}$
, which is defined as \( \frac{d \mathcal{L}(\textbf{x}^{(\ell)})}{d \textbf{x}^{(\ell)}} \). This gradient is expressed as a row vector. With a learning rate of \( \gamma \), a standard gradient descent update policy for transform order pair can be formulated as follows:

\begin{equation}
\begin{pmatrix}
\alpha_{\mathrm{next}} \\
\beta_{\mathrm{next}}
\end{pmatrix}= 
\begin{pmatrix}
\alpha_{\mathrm{current}} \\
\beta_{\mathrm{current}}
\end{pmatrix}
- \gamma
\begin{pmatrix}
\nabla_{\mathbf{x}}^{(\ell)} \mathcal{L} \cdot \frac{\partial \mathbf{x}^{(\ell)}}{\partial \alpha} \\
\nabla_{\mathbf{x}}^{(\ell)} \mathcal{L} \cdot \frac{\partial \mathbf{x}^{(\ell)}}{\partial \beta}
\end{pmatrix}.
\end{equation}

\begin{equation}
\begin{pmatrix}
\frac{\partial \mathbf{x}^{(\ell)}}{\partial \alpha} \\
\frac{\partial \mathbf{x}^{(\ell)}}{\partial \beta}
\end{pmatrix}
= 
\begin{pmatrix}
\dot{\varphi} \left( \mathbf{F}_{J}^{\alpha, \beta} \mathbf{x}^{(\ell-1)} \right) \odot \left( \mathbf{F}_{J}^{\dot{\alpha}, \beta} \mathbf{x}^{(\ell-1)} \right) \\
\dot{\varphi} \left( \mathbf{F}_{J}^{\alpha, \beta} \mathbf{x}^{(\ell-1)} \right) \odot \left( \mathbf{F}_{J}^{\alpha, \dot{\beta}} \mathbf{x}^{(\ell-1)} \right)
\end{pmatrix}.
\end{equation}

\begin{equation}
{\textbf{F}}_{J}^{\dot\alpha, \beta} = {{\textbf{F}}^\beta} \otimes \dot{\textbf{F}}_{G}^\alpha,\quad{\textbf{F}}_{J}^{\alpha, \dot\beta} = \dot{\textbf{F}}^\beta \otimes {\textbf{F}}_{G}^\alpha.
\end{equation}

\section{Numerical Experiments and Results}
In this section, we will design simulation experiments using the proposed learnable JFRFT to evaluate the accuracy and the denoising performance of our method. Through these simulations, we aim to validate the effectiveness of the learnable JFRFT in real-world scenarios and ensure that the method produces the expected results.

\subsection{Transform Learning}
Given the graph signal transformed by the JFRFT, our goal is to learn the graph fractional order parameter \( \alpha \) and the fractional order parameter \( \beta \) that characterize the graph signal. Therefore, we randomly select a time-varying graph signal \( \mathbf{X} \in \mathbb{R}^{20 \times 6} \)and assign its values using a uniform distribution. We then apply the JFRFT using the original fractional orders \( \alpha_{\text{orig}} \) and \( \beta_{\text{orig}} \) to transform the signal. The target signal is then obtained as \( \mathbf{Y} = \mathbf{F}_G^{\alpha_{\text{orig}}} \mathbf{X}\left( \mathbf{F}^{\beta_{\text{orig}}} \right)^T \). 

Next, we randomly generate the adjacency matrix \( \mathbf{A} \in \mathbb{R}^{20 \times 20} \) for a weighted directed graph, where the elements are independently drawn from a uniform distribution. 

Based on this setup, we introduce a multi-layer JFRFT network, where the transform order pair of the \( \ell \)-th layer is denoted by \( \alpha_{\ell} \) and \( \beta_{\ell} \). The entire network can be formulated as follows:

\begin{equation}
\widehat{\textbf{Y}} = \left( \prod_{\ell=1}^L \textbf{F}_{G}^{\alpha_{\ell}} \right) \mathbf{X} \left( \prod_{\ell=1}^L \left(\textbf{F}^{\beta_{\ell}}\right)^T \right).
\end{equation}

\begin{equation}
\mathcal{L}_{\text{MSE}} = \frac{1}{NT} \left\| \textbf{Y} - \widehat{\textbf{Y}} \right\|_F^2.
\end{equation}

Here, \( \widehat{\textbf{Y}} \) denotes the estimated output of the network and \( \mathcal{L}_{\text{MSE}} \) represents the mean squared error loss function. Notably, the trainable parameters of the network include both \( \alpha_{\ell} \) and \( \beta_{\ell} \). The network is trained using the Adam optimizer for 1200 epochs with a learning rate of \( 1 \times 10^{-3} \). Experiments are conducted for \( L \in \{1, 2, 3\} \), with the initial fractional orders \( (\alpha_{\text{orig}}, \beta_{\text{orig}}) \) set to \( (0.55, 0.45) \) and \( (1.55, 1.45) \). The entire training process takes only a few seconds, with the results summarized in Table~\ref{tab:transform}. The initial transform order pairs are assigned in the first epoch, and in all cases, the sum of the learned transform order pairs converges to the original values \( \alpha_{\text{orig}} \) and \( \beta_{\text{orig}} \). The complete learning procedure is detailed in the Algorithm 1. This experiment demonstrates that the proposed learnable JFRFT can learn both the graph fractional order and the fractional order. Furthermore, their orders are shown to exhibit the index additivity property.

\begin{algorithm}
\caption{Transform learning}
\begin{algorithmic}[1]
\State \textbf{Input:} $\textbf{Y} = \mathbf{F}_G^{\alpha_{\text{orig}}} \mathbf{X}\left( \mathbf{F}^{\beta_{\text{orig}}}\right)^T \in \mathbb{C}^{N \times T}$, learning rate $\gamma$, training epochs
\State \textbf{Output:} Trained fractional orders $\sum_{i=1}^L \alpha_i$, $\sum_{i=1}^L \beta_i$

\State \textbf{Initialization:}
\State Initialize signal $\mathbf{X}$ randomly from uniform distribution
\State Initialize the number of layers $L$
\For{each layer $\ell = 1$ to $L$}
    \State Initialize $\alpha_{\ell}$ and $\beta_{\ell}$  
\EndFor

\State \textbf{Step 1: Construct Operators}
\For{layer $\ell = 1$ to $L$}
    \State Compute the fractional graph transform matrix $\mathbf{F}_G^{\alpha_{\ell}}$ for layer $\ell$
    \State Compute the fractional Fourier transform matrix $\mathbf{F}^{\beta_{\ell}}$ for layer $\ell$
\EndFor

\State \textbf{Step 2: Forward Propagation}
\State Compute the output of the network as: $\widehat{\textbf{Y}} = \left( \prod_{\ell=1}^L \mathbf{F}_G^{\alpha_{\ell}} \right) \mathbf{X} \left( \prod_{\ell=1}^L \left(\mathbf{F}^{\beta_{\ell}}\right)^T \right)$

\State \textbf{Step 3: Loss Calculation and Backpropagation}
\State Compute the loss function: $L = \frac{1}{N T} \left\| \widehat{\textbf{Y}} - \textbf{Y} \right\|_F^2$
\State Use Adam optimizer to update the parameters $\alpha_{\ell}$ and $\beta_{\ell}$ for each layer

\State \textbf{Step 4: Repeat Training Process}
\For{epoch = 1 to training epochs}
    \State Apply forward propagation to compute the output $\widehat{\textbf{Y}}$
    \State Compute MSE loss and update $\alpha_{\ell}$, $\beta_{\ell}$ for all layers
\EndFor

\State \textbf{Step 5: Output Results}
\State Compute the sum of the fractional orders across all layers: $\sum_{i=1}^L \alpha_i$ and $\sum_{i=1}^L \beta_i$

\end{algorithmic}
\end{algorithm}

\begin{sidewaystable}[!htbp]
\caption{Convergence of the transform order pair in transform learning experiment}
\label{tab:transform}
\resizebox{\textwidth}{!}{
\renewcommand{\arraystretch}{2.5}
\begin{tabular}{cccccccccccccccc}
\toprule
\multirow{2}{*}{original orders} & \multirow{2}{*}{Epoch} & \multicolumn{2}{c}{L=1} & \multicolumn{4}{c}{L=2} & \multicolumn{5}{c}{L=3} \\ \cline{3-13} 
&& \(\left(\alpha_1, \beta_1\right)\) & loss & \(\left(\alpha_1, \beta_1\right)\) & \(\left(\alpha_2, \beta_2\right)\) & sum & loss & \(\left(\alpha_1, \beta_1\right)\) & \(\left(\alpha_2, \beta_2\right)\) & \(\left(\alpha_3, \beta_3\right)\) & sum & loss \\ \midrule
\multirow{5}{*}{(0.45,0.55)} 
& 1 & \(\left(0.0000, 0.0000\right)\) & 0.792 & \(\left(0.0000, 0.0000\right)\) & \(\left(1.0000, 1.0000\right)\) & \(\left(1.0000, 1.0000\right)\) & 1.36 & \(\left(0.0000, 0.0000\right)\) & \(\left(0.2500, 0.2500\right)\) & \(\left(0.5000, 0.5000\right)\) & \(\left(0.7500, 0.7500\right)\) & 0.455 \\
& 300 & \(\left(0.3248, 0.4121\right)\) & \(1.48 \times 10^{-1}\) & \(\left(-0.2176, -0.2252\right)\) & \(\left(0.7824, 0.7748\right)\) & \(\left(0.5647, 0.5496\right)\) & \(2.59 \times 10^{-2}\) & \(\left(-0.0999, -0.0667\right)\) & \(\left(0.1501, 0.1833\right)\) & \(\left(0.4001, 0.4333\right)\) & \(\left(0.4503, 0.5500\right)\) & \(2.34 \times 10^{-7}\) \\
& 600 & \(\left(0.4478, 0.5468\right)\) & \(8.27 \times 10^{-5}\) & \(\left(-0.2692, -0.2253\right)\) & \(\left(0.7308, 0.7747\right)\) & \(\left(0.4617, 0.5494\right)\) & \(2.58 \times 10^{-4}\) & \(\left(-0.1000, -0.0667\right)\) & \(\left(0.1500, 0.1833\right)\) & \(\left(0.4000, 0.4333\right)\) & \(\left(\mathbf{0.4500, 0.5500}\right)\) & \(1.22 \times 10^{-12}\) \\
& 900 & \(\left(\mathbf{0.4500, 0.5500}\right)\) & \(6.90 \times 10^{-10}\) & \(\left(-0.2748, -0.2250\right)\) & \(\left(0.7252, 0.7750\right)\) & \(\left(0.4505, 0.5500\right)\) & \(4.19 \times 10^{-7}\) & \(\left(-0.1000, -0.0667\right)\) & \(\left(0.1500, 0.1833\right)\) & \(\left(0.4000, 0.4333\right)\) & \(\left(\mathbf{0.4500, 0.5500}\right)\) & \(1.18 \times 10^{-12}\) \\
& 1200 & \(\left(\mathbf{0.4500, 0.5500}\right)\) & \(8.26 \times 10^{-12}\) & \(\left(-0.2750,-0.2250\right)\) & \(\left(0.7250, 0.7750\right)\) & \(\left(\mathbf{0.4500, 0.5500}\right)\) & \(8.35 \times 10^{-11}\) & \(\left(-0.1000,-0.0667\right)\) & \(\left(0.1500, 0.1833\right)\) & \(\left(0.4000, 0.4333\right)\) & \(\left(\mathbf{0.4500, 0.5500}\right)\) & \(1.16 \times 10^{-12}\) \\
\midrule
\multirow{5}{*}{(1.45,1.55)}  
& 1 & \(\left(1.0000, 1.0000\right)\) & 3.31 & \(\left(0.0000, 0.0000\right)\) & \(\left(1.0000,1.0000\right)\) & \(\left(1.0000, 1.0000\right)\) & 3.31 & \(\left(0.0000, 0.0000\right)\) & \(\left(0.7500, 0.7500\right)\) & \(\left(1.2500, 1.2500\right)\) & \(\left(2.0000, 2.0000\right)\) & 10.6 \\
& 300 & \(\left(1.2568, 1.4387\right)\) & \(7.22 \times 10^{-1}\) & \(\left(0.2235, 0.2751\right)\) & \(\left(1.2235, 1.2751\right)\) & \(\left(1.4471, 1.5503\right)\) & \(9.31 \times 10^{-5}\) & \(\left(-0.1588, -0.1506\right)\) & \(\left(0.5912, 0.5994\right)\) & \(\left(1.0912, 1.0994\right)\) & \(\left(1.5236, 1.5481\right)\) & \(5.68 \times 10^{-2}\) \\
& 600 & \(\left(1.4469, 1.5500\right)\) & \(1.00 \times 10^{-4}\) & \(\left(0.2250, 0.2750\right)\) & \(\left(1.2250, 1.2750\right)\) & \(\left(\mathbf{1.4500, 1.5500}\right)\) & \(1.01 \times 10^{-11}\) & \(\left(-0.1819, -0.1500\right)\) & \(\left(0.5681, 0.6000\right)\) & \(\left(1.0681, 1.1000\right)\) & \(\left(1.4543, 1.5499\right)\) & \(1.85 \times 10^{-4}\) \\
& 900 & \(\left(\mathbf{1.4500, 1.5500}\right)\) & \(4.01 \times 10^{-10}\) & \(\left(0.2250, 0.2750\right)\) & \(\left(1.2250, 1.2750\right)\) & \(\left(\mathbf{1.4500, 1.5500}\right)\) & \(6.50 \times 10^{-12}\) & \(\left(-0.1833, -0.1500\right)\) & \(\left(0.5667, 0.6000\right)\) & \(\left(1.0667, 1.1000\right)\) & \(\left(1.4501, 1.5500\right)\) & \(5.62 \times 10^{-8}\) \\
& 1200 & \(\left(\mathbf{1.4500, 1.5500}\right)\) & \(1.47 \times 10^{-10}\) & \(\left(0.2250, 0.2750\right)\) & \(\left(1.2250, 1.2750\right)\) & \(\left(\mathbf{1.4500, 1.5500}\right)\) & \(4.46 \times 10^{-12}\) & \(\left(-0.1833, -0.1500\right)\) & \(\left(0.5667, 0.6000\right)\) & \(\left(1.0667, 1.1000\right)\) & \(\left(\mathbf{1.4500, 1.5500}\right)\) & \(6.64 \times 10^{-11}\) \\
\bottomrule  
\end{tabular}}
\end{sidewaystable}

\clearpage

\subsection{Optimal Transform Order Pair Learning}
\subsubsection{Denoising Task}
The Wiener filtering in the JFRFT domain was considered for time-varying graph signals, as discussed in \cite{ref23}. Let \(\textbf{X} \in \mathbb{C}^{N \times T}\) represent the time-varying graph signal and \(\textbf{N} \in \mathbb{C}^{N \times T}\) represent the noise. We define \(\mathbf{x} = \operatorname{vec}(\mathbf{X})\) and \(\mathbf{n} = \operatorname{vec}(\mathbf{N})\), known expectations given by \( E\{\textbf{x}\textbf{x}^H\} \), \( E\{\textbf{n}\textbf{n}^H\} \), \( E\{\textbf{x}\textbf{n}^H\} \) and \( E\{\textbf{n}\textbf{x}^H\} \). The formulation for the received time-varying graph signal is then expressed as equation (23) with known arbitrary transforms $\textbf{G}_T,\textbf{G}_G$.

\begin{equation}
\textbf{Y} = \textbf{G}_G \textbf{X} \textbf{G}_T + \textbf{N}, \ \text{and} \quad   \textbf{y} = (\textbf{G}_T^T \otimes \textbf{G}_G) \textbf{x} + \textbf{n}.
\end{equation}

The optimal Wiener filtering problem in the JFRFT domain can be formulated as follows:

\begin{equation}
\min_{\mathbf{H}_J \in \mathcal{D}} \mathbb{E}\left\{ \left\| \mathbf{F}_J^{-\alpha,-\beta} \mathbf{H}_J \mathbf{F}_J^{\alpha,\beta} \mathbf{y} - \mathbf{x} \right\|_2^2 \right\}
\end{equation}

Hence, we define \( \textbf{W}_m =\textbf{w}_m \tilde{\textbf{w}}^T_m \), where \( \tilde{\textbf{w}}_m \) is the \( m \)-th row of \( \textbf{F}^{\alpha,\beta}_J \) and the \( m \)-th column of \( \textbf{F}^{-\alpha,-\beta}_J \), for \( m \in \{1, \dots, NT\} \). Equation (24) is shown to be a convex optimization problem. Let \( \textbf{S} = [\textbf{s}_1,\textbf{s}_2 \cdots \textbf{s}_{NT}] \) with \( \textbf{s}_k = \textbf{W}_k \textbf{y} \). Then, the optimal coefficients can be obtained by solving \( \textbf{T} \textbf{h}^{(\text{opt})} = \textbf{q} \), where \( \textbf{h} = \text{diag}(\textbf{H}_J) \), \( \textbf{T} = E\{ \textbf{S}^H \textbf{S} \} \), \( \textbf{q} = E\{ \textbf{S}^H \textbf{x} \} \).

In our work, we consider a time-varying graph signal \( \textbf{X} \in \mathbb{C}^{N \times MT} \). Instead of reshaping the signal into a single column vector, we divide it into \( M \) blocks by columns. Each block is represented as a matrix \( \textbf{X}_i \in \mathbb{C}^{N \times T} \) for \( i = 1, 2, \dots, M \). We then reshape each matrix into a column vector \( \textbf{x}_i \in \mathbb{C}^{NT \times 1} \). Finally, we compile these vectors into a set denoted as \( {\textbf{x}} = \{ \textbf{x}_1, \textbf{x}_2, \dots, \textbf{x}_M \} \). To generate the optimal filter as ${\textbf{H}}_{\text{opt}}$, we use the mean of the auto-correlation and cross-correlation. For example, the auto-correlation of the graph signal is calculated as \( \frac{1}{M} \sum_{i=1}^{M} E\left\{ \textbf{x}_i {\textbf{x}_i}^H \right\} \). In the case where \( M = 1 \), this reduces to the classical JFRFT-based Wiener filter. Consequently, we obtain the estimated graph signal as follows:

\begin{equation}
\hat{\textbf{x}} = \textbf{F}_J^{-\alpha,-\beta} \textbf{H}_{\text{opt}} {\textbf{F}}_J^{\alpha,\beta} \left( ({\textbf{G}}_{T}^{T} \otimes \textbf{G}_G) \textbf{x} + \textbf{n} \right).
\end{equation}

In the denoising experiment described below, let $\textbf{G}_{T}^{T}=\textbf{G}_G=\textbf{I}$. we also utilize the learnable JFRFT from Section 3 to learn the optimal transform order pair and filter coefficients for denoising. We assess the effectiveness of the denoising process using the following signal-to-noise ratio (SNR):

\begin{equation}
SNR = 10 \log_{10} \left( \frac{\|\textbf{X}\|_F^2}{\|\textbf{X} - \hat{\textbf{X}}\|_F^2} \right). 
\end{equation}

\subsubsection{Denoising Results}
Considering a \( K \)-\( L \) bandlimited graph signal defined in Section 3, we introduce the concept of the `high-frequency' graph noise that arises in the JFRFT domain after transforming the graph signal. For clarity, we illustrate in Figure 1 below that the graph signal and noise are completely separated after applying the JFRFT. The green-highlighted section represents the signal, while the red-highlighted section indicates the noise. Under this condition, the JFRFT-based filtering method can effectively remove the noise.

\begin{figure}[!htbp]
    \centering
    \includegraphics[width=0.6\textwidth]{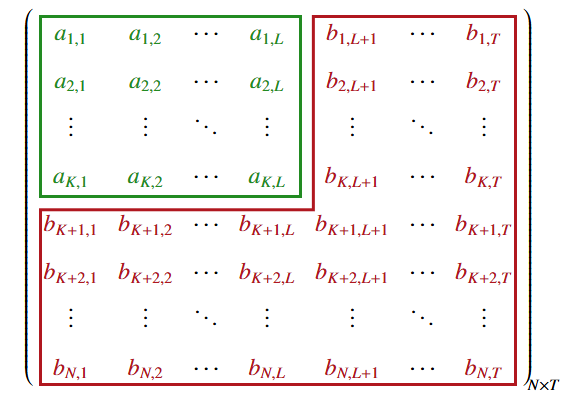} 
    \caption{Separable graph signal and noise in \(\left(\alpha,\beta\right)\) domain}  
    \label{fig:1}  
\end{figure}

 However, when overlap occurs, it is not possible to completely eliminate the noise. In Figure 2, we demonstrate the \( Q \)-\( P \) bandlimted graph signal contains overlapping regions with `high-frequency' noise after applying the JFRFT. The green-highlighted section represents the part of the graph signal that does not overlap with the noise. The purple-highlighted section represents the overlap between the graph signal and noise. The red-highlighted section represents the part of the graph noise that does not overlap with the signal.
\clearpage

\begin{figure}[!htbp]
    \centering
    \includegraphics[width=1\textwidth]{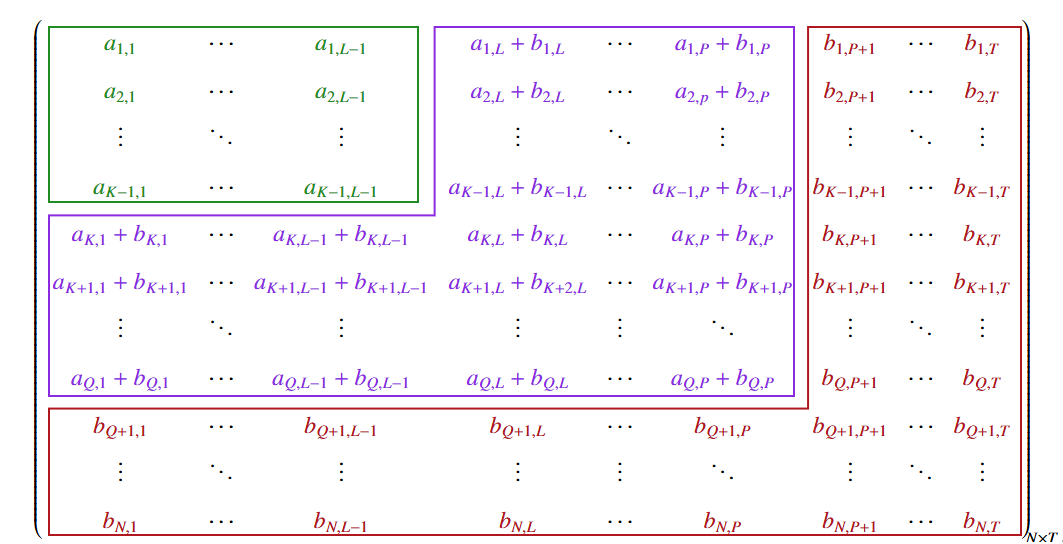}  
    \caption{Overlapping graph signal and noise in \(\left(\alpha,\beta\right)\) domain}  
    \label{fig:2}  
\end{figure}

Despite the challenges, it is possible to determine an optimal transform order pair and filter coefficients that minimize noise in a noisy graph signal. To demonstrate this experimentally, we utilize the learnable JFRFT to optimize these parameters. We synthesize \( Q \)-\( P \) bandlimited graph signals and noise using the real-world Sea Surface Temperature (SST) dataset. We add the `high-frequency' noise into the graph signal within the JFRFT domain. This noise follows a zero-mean Gaussian distribution with a standard deviation of \( \sigma \). After applying the inverse transformation, we obtain the graph noise in the vertex domain.

In the following experiments, we will consider two scenarios: one in which the graph signal and noise in the $(0.55, 0.45)$ domain are non-overlapping and another in which they overlap. We divide the time-varying graph signal \( \mathbf{X} \in \mathbb{C}^{6 \times 36} \) into six blocks, each block matrix remains \( Q \)-\( P \) bandlimited after transformation. For simplicity, we set \( Q =P=4 \) and \( K =L\), assuming that the overlapping regions are the same size. Similarly, we divide the noise \( \mathbf{N} \in \mathbb{C}^{6 \times 36} \) into six blocks, with each transformed noise matrix exhibiting high-frequency components. The overlapping region consists of 0, 2, or 4 components, i.e., \( P - L + 1 \in \{ 0, 2, 4 \} \). The fixed ideal low-pass diagonal graph filter is 
$\mathbf{H}_{\text{fixed}} \triangleq \operatorname{diag} \left( \underbrace{1, 1, 1, 1, 0, 0 }_{\text{repeated 4 times}}\dots,\textbf{0}_{12\times1} \right)$. The learnable diagonal filter is 
\(\mathbf{H}_{\text{learn}} \triangleq \operatorname{diag} \left(  h_1, \dots, h_{36}  \right) \), 
where each entry \( h_i \) is a trainable parameter. The optimal filter ${\textbf{H}}_{\text{opt}}$ is defined above. 
We will experiment using these filters and have the estimated graph signal:

\begin{equation}
\hat{\mathbf{x}} = \mathbf{F}_J^{-\alpha,-\beta} \mathbf{H} \mathbf{F}_J^{\alpha,\beta} \mathbf{y},
\ \text{for} \ \mathbf{H} \in \{ \mathbf{H}_{\text{fixed}}, \mathbf{H}_{\text{learn}}, \mathbf{H}_{\text{opt}} \}.
\end{equation}

Subsequently, the transform order pair (\(\alpha\), \(\beta\)) and the filter coefficients $h_i$ are learned by minimizing the MSE loss function. The coefficients of \( \textbf{H}_{\text{learn}} \) are initialized to 1 and the transform orders are set to (0.1, 0.1). The overall denoising procedure is outlined in Algorithm 2. To compare with the learnable GFRFT, we also initialize the transform order of the GFRFT to 0.1. Finally, we train the network for 10,000 epochs with a learning rate of $5 \times 10^{-3}$. The entire process takes less than a minute. We also apply the grid search strategy for GFRFT-based filtering and JFRFT-based filtering, where the order ranges from $-2$ to 2 with a step size of 0.01. This method takes significantly more time than the learnable GFRFT and learnable JFRFT. The denoising results are summarized in Table~\ref{tab:sy}. We observe that the proposed learnable JFRFT-based filtering achieves complete noise separation, with the transform orders converging to the original values when \(\text{overlap} = 0\), where the MSE error is in the range of \(10^{-14}\) for \(\textbf{H}_{\text{fixed}}\) and \(10^{-9}\) for \(\textbf{H}_{\text{learn}}\).
 For the case where the overlap is not equal to zero, we observe that the proposed learnable JFRFT-based filtering outperforms other filtering methods in terms of noise separation.

\begin{algorithm}[!htbp]
\caption{Learnable JFRFT denoising experiments for synthetic data}
\begin{algorithmic}[1]
\State \textbf{Input:} $\textbf{Y} = \textbf{G}_G \textbf{X} \textbf{G}_T + \textbf{N} \in \mathbb{C}^{N \times MT}$, learning rate $\gamma$, training epochs, $\textbf{H}_\text{fixed} \in \mathbb{C}^{N \times N}$ 
\State \textbf{Output:} $\alpha_{out}$, $\beta_{out}$, $\textbf{H}_{out}$, $SNR$

\State \textbf{Initialization:}
\State Initialize $\alpha_0 = 0.1$
\State Initialize $\beta_0 = 0.1$
\State Initialize $\textbf{H}_0 = \textbf{I}$

\State \textbf{Step 1: Vectorized reorganization}
\State Split $Y$ into blocks: $\textbf{Y} = [\textbf{Y}_T, \textbf{Y}_{2T}, \dots, \textbf{Y}_{MT}] \in \mathbb{C}^{N \times MT}$
\State For each block, reshape each column: $\textbf{y} = [\textbf{y}_T, \textbf{y}_{2T}, \dots, \textbf{y}_{MT}] \in \mathbb{C}^{NT \times M}$

\State \textbf{Step 2: Construct Operator}
    \State Compute ${\mathbf{F}}^{\beta_0} \in \mathbb{C}^{T \times T}$ \text{and} ${\mathbf{F}}_G^{\alpha_0} \in \mathbb{C}^{N \times N}$
    \State Compute $\textbf{F}_J^{\alpha_0, \beta_0} \triangleq \mathbf{F}^{\beta_0} \otimes \mathbf{F}_G^{\alpha_0} \in \mathbb{C}^{NT \times NT}$

\State \textbf{Step 3: Forward propagation}
\If{$\textbf{H}_\text{fixed}$ is used}
    \State Compute $\textbf{y}_0 = \textbf{F}_J^{-\alpha_0, -\beta_0} \textbf{H}_\text{fixed} \textbf{F}_J^{\alpha_0, \beta_0} \textbf{y} \in \mathbb{C}^{NT \times M}$
\Else
    \State Compute $\textbf{y}_0 = \textbf{F}_J^{-\alpha_0, -\beta_0} \textbf{H}_0 \textbf{F}_J^{\alpha_0, \beta_0} \textbf{y} \in \mathbb{C}^{NT \times M}$
\EndIf
\State Reshape $\textbf{y}_0$ back to matrix form: $\textbf{Y}_0 \in \mathbb{C}^{N \times MT}$

\State \textbf{Step 4: Loss calculation and backpropagation}
\State Compute the loss function
\State Use Adam optimizer to update parameters: $\alpha_1, \beta_1, \textbf{H}_1$ (if not using $\textbf{H}_\text{fixed}$)

\State \textbf{Step 5: Repeat training process}
\For{epoch = 1 to training epochs}
    \If{$\textbf{H}_\text{fixed}$ is used}
        \State Only update $\alpha$ and $\beta$, keep $\textbf{H}_\text{fixed}$ fixed
        \State Compute new $\textbf{y}_0 = \textbf{F}_J^{-\alpha_1, -\beta_1} \textbf{H}_\text{fixed} \textbf{F}_J^{\alpha_1, \beta_1} \textbf{y}$
    \Else
        \State Compute new $\textbf{y}_0 = \textbf{F}_J^{-\alpha_1, -\beta_1} \textbf{H}_1 \textbf{F}_J^{\alpha_1, \beta_1} \textbf{y}$
        \State Update $\alpha_1$, $\beta_1$, $\textbf{H}_1$ using Adam optimizer
    \EndIf
    \State Compute new $\textbf{Y}_0$ and update $\alpha_1$, $\beta_1$, $\textbf{H}_1$ (if not using $\textbf{H}_\text{fixed}$)
\EndFor

\State \textbf{Step 6: Compute SNR}
\State Compute SNR from the output of the current experiment: $SNR = 10 \log_{10} \left( \frac{\|\textbf{X}\|_F^2}{\|\textbf{X} - \textbf{Y}_0\|_F^2} \right)$

\State \textbf{Store result:}
\State $\alpha_{out}, \beta_{out}, \textbf{H}_{out}$ are the trained parameters obtained from this experiment.

\end{algorithmic}
\end{algorithm}

\clearpage 

\begin{sidewaystable}[!htbp]
\caption{SNR results for transform order pair learning experiment on the synthetic dataset}
\label{tab:sy}
\resizebox{\textwidth}{!}{
\renewcommand{\arraystretch}{3.5}
\begin{tabular}{lccccccccc}
\toprule
\multicolumn{1}{c}{\multirow{3}{*}{Method}} & \multicolumn{3}{c}{overlap = 0}& \multicolumn{3}{c}{overlap = 2} & \multicolumn{3}{c}{overlap = 4} \\ \cline{2-10} 
\multicolumn{1}{c}{}& \multicolumn{1}{c}{$\sigma = 0.2$} & \multicolumn{1}{c}{$\sigma = 0.3$} & \multicolumn{1}{c}{$\sigma = 0.6$} & \multicolumn{1}{c}{$\sigma = 0.2$} & \multicolumn{1}{c}{$\sigma = 0.3$} & \multicolumn{1}{c}{$\sigma = 0.6$} & \multicolumn{1}{c}{$\sigma = 0.2$} & \multicolumn{1}{c}{$\sigma = 0.3$} & \multicolumn{1}{c}{$\sigma = 0.6$}  \\ 
\multicolumn{1}{c}{}& 11.07& 7.54& 1.52& 9.23& 5.71& $-$0.32& 8.80& 5.28& $-$0.75\\ \hline
GFT-H\textsubscript{learn}& 12.50& 9.84& 6.01& 11.01& 8.43& 4.68& 10.49& 7.88& 4.12\\
GFRFT-H\textsubscript{opt}&(0.55)17.18& (0.54)13.94& (0.53)8.82& (0.54)13.83& (0.54)10.75& (0.55)6.23&(0.54)12.86& (0.53)9.85& (0.52)5.40\\
GFRFT-H\textsubscript{learn}& (0.546)17.19& (0.544)13.95& (0.534)8.83& (0.543)13.92 & (0.537)10.83 & (0.543)6.30& (0.54)12.99 & (0.528)9.97 & (0.523)5.51 \\
JFT-H\textsubscript{fixed}& 4.97& 4.59& 2.85& 4.70& 3.99& 1.40 & 4.61& 3.82 & 1.04 \\
JFT-H\textsubscript{learn}& 14.95& 12.32& 8.36& 13.49& 10.96& 7.52& 12.73& 10.16 & 6.59\\
JFRFT-H\textsubscript{fixed}& \textbf{(0.55,0.45)130.92}& \textbf{(0.55,0.45)130.69}& \textbf{(0.55,0.45)129.32}& (0.553,0.449)13.85& (0.557,0.447)10.33& (0.578,0.444)4.32& (0.554,0.450)12.70& (0.558,0.449)9.18& (0.583,0.449)3.17\\
JFRFT-H\textsubscript{opt}& (0.55,0.45)129.87& (0.55,0.45)129.79& (0.55,0.45)127.47& (0.54,0.46)16.58& (0.54,0.47)13.78& (0.57,0.48)9.52& (0.54,0.46)14.70& (0.53,0.46)11.72& (0.53,0.44)7.07\\
JFRFT-H\textsubscript{learn}& (0.550,0.450)103.01& (0.550,0.449)99.03& (0.549,0.449)81.94& \textbf{(0.546,0.460)17.60}& \textbf{(0.547,0.470)14.75}& \textbf{(0.605,0.488)10.46}& \textbf{(0.545,0.459)15.46}& \textbf{(0.542,0.467)12.42}& \textbf{(0.556,0.440)7.65}\\
\bottomrule
\end{tabular}}
\end{sidewaystable}

\clearpage 
At the same time, Gaussian white noise of varying intensities is added to the real-world datasets of SST, Particulate Matter 2.5 (PM-2.5) and COVID-19 global (COVID) to evaluate the denoising performance across different graph structures. Five types of shift operators are considered: adjacency, Laplacian, row-normalized adjacency, symmetric-normalized adjacency and normalized Laplacian. We select 10 vertices and 60 time points in the dataset and initialize all the coefficients of $\textbf{H}_{\text{learn}}$ to 1. In the case of the learnable JFRFT, $\alpha$ is randomly selected from the interval [$-2$, 2] and $\beta$ is also randomly chosen from the same interval. The two randomly selected values form the initialized fractional order pair $(\alpha, \beta)$. The network is then trained for 10,000 epochs with a learning rate of \( 5 \times 10^{-3} \), after which a new transform order pair is randomly selected. This process is repeated over 20 independent runs to obtain the optimal result for each graph structure. The full experimental workflow is summarized in Algorithm 3. Similarly, for the learnable GFRFT, we randomly select a value from the interval [$-2$, 2] for the fractional order and perform 20 experiments to obtain the optimal result for this graph structure. We also apply the search method: a subscript of 1 indicates the use of the grid search method, while a subscript of 2 denotes the use of the learnable method. 

We compare JFRFT-based filtering with GFRFT-based filtering and ARMA$_{K}$ graph filters \cite{ref39}, where $K \in \{3, 4, 5\}$, with the optimal cutoff frequency $\lambda_c$ determined by sweeping $\lambda_c \in [\lambda_{\text{min}}, \lambda_{\text{max}}]$ in steps of 0.01. Additionally, we compare both the Median$_1$ and Median$_2$ filtering approaches \cite{ref40}. Finally, we use three classic graph neural network (GNN) models for denoising, including graph convolutional network (GCN) \cite{ref41}, graph attention network (GAT) \cite{ref42}, and Chebyshev graph convolutional network (ChebNet) \cite{ref43}. We select data from the dataset consisting of 10 vertices and 300 time points, and split it into training and test sets in an 8:2 ratio, where the test set contains data from 60 time points that need to be denoised. The training set is further divided into training and validation sets in an 8:2 ratio. The entire training process uses the Adam optimizer, and the model is trained under different hyperparameter configurations, with the best-performing test result selected at the end.

The results presented in Table~\ref{tab:SST}, Table~\ref{tab:PM2.5} and Table~\ref{tab:COVID}
, corresponding to different datasets, demonstrate that the proposed learnable JFRFT, based on a gradient backpropagation mechanism, consistently outperforms other methods in terms of filtering performance. In Table~\ref{tab:Comparsion of different methods}, we present a comparison between the proposed method and the baseline methods in terms of parameter count, feature characterization, computational complexity, the need for prior information, and the requirement for eigen decomposition. It can be seen that the proposed method incorporates the spatiotemporal features of time-varying graph signals, and compared to the traditional JFRFT grid search strategy, it significantly reduces computational complexity.

\begin{algorithm}
\caption{Learnable JFRFT denoising experiments for real data}
\begin{algorithmic}[1]
\State \textbf{Input:} $\textbf{Y} = \textbf{G}_G \textbf{X} \textbf{G}_T + \textbf{N} \in \mathbb{C}^{N \times MT}$, learning rate $\gamma$, training epochs
\State \textbf{Output:} $\alpha_{out}$, $\beta_{out}$, $\textbf{H}_{out}$, $SNR_{max}$

\State \textbf{Initialization:}
\State Randomly initialize $\alpha_0 \in [-2, 2]$
\State Randomly initialize $\beta_0 \in [-2, 2]$
\State Initialize $\textbf{H}_0 = \textbf{I}$ 

\For{experiment = 1 to 20}
    \State \textbf{Step 1: Vectorized reorganization}
    \State Split $\textbf{Y}$ into blocks: $\textbf{Y} = [\textbf{Y}_T, \textbf{Y}_{2T}, \dots, \textbf{Y}_{MT}] \in \mathbb{C}^{N \times MT}$
    \State For each block, reshape each column: $\textbf{y} = [\textbf{y}_T, \textbf{y}_{2T}, \dots, \textbf{y}_{MT}] \in \mathbb{C}^{NT \times M}$

    \State \textbf{Step 2: Construct Operator}
    \State Compute ${\mathbf{F}}^{\beta_0} \in \mathbb{C}^{T \times T}$ \text{and} ${\mathbf{F}}_G^{\alpha_0} \in \mathbb{C}^{N \times N}$
\State Compute $\mathbf{F}_J^{\alpha_0, \beta_0} \triangleq \mathbf{F}^{\beta_0} \otimes \mathbf{F}_G^{\alpha_0} \in \mathbb{C}^{NT \times NT}$

    \State \textbf{Step 3: Forward propagation}
    \State Compute $\textbf{y}_0 = \textbf{F}_J^{-\alpha_0, -\beta_0} \textbf{H}_0 \textbf{F}_J^{\alpha_0, \beta_0} \textbf{y} \in \mathbb{C}^{NT \times M}$
    \State Reshape $\textbf{y}_0$ back to matrix form: $\textbf{Y}_0 \in \mathbb{C}^{N \times MT}$

    \State \textbf{Step 4: Loss calculation and backpropagation}
    \State Compute the loss function
    \State Use Adam optimizer to update parameters: $\alpha_1, \beta_1, \textbf{H}_1$

    \State \textbf{Step 5: Repeat training process}
    \For{epoch = 1 to training epochs}
        \State Compute new $\textbf{y}_0 = \textbf{F}_J^{-\alpha_1, -\beta_1} \textbf{H}_1 \textbf{F}_J^{\alpha_1, \beta_1} \textbf{y}$
        \State Compute new $\textbf{Y}_0$ and update $\alpha_1$, $\beta_1$, $\textbf{H}_1$ using Adam optimizer
    \EndFor

    \State \textbf{Step 6: Compute SNR}
    \State Compute SNR from the output of the current experiment: $SNR = 10 \log_{10} \left( \frac{\|\textbf{X}\|_F^2}{\|\textbf{X} - \textbf{Y}_0\|_F^2} \right)$
\EndFor
\State \textbf{Store result:}
\State Store maximum SNR from 20 experiments: $SNR_{max} = \max(SNR_1, SNR_2, \dots, SNR_{20})$
\State $\alpha_{out}, \beta_{out}, \textbf{H}_{out}$ are the trained parameters obtained from the experiment where $SNR_{max}$ was achieved.
\end{algorithmic}
\end{algorithm}

\clearpage
\begin{sidewaystable}[!htbp]

\centering

\footnotesize
\setlength{\tabcolsep}{18pt}
\renewcommand{\arraystretch}{1.4}
\caption{SNR results for transform order pair learning experiment for SST}
\label{tab:SST}
\begin{tabular}{cccccccccc}
\toprule
& \multicolumn{3}{c}{2-NN} & \multicolumn{3}{c}{3-NN} & \multicolumn{3}{c}{5-NN} \\ \cline{2-10}
& \multicolumn{1}{c}{$\sigma = 0.2$}  & \multicolumn{1}{c}{$\sigma = 0.25$} &  \multicolumn{1}{c}{$\sigma = 0.3$}                 
& \multicolumn{1}{c}{$\sigma = 0.2$}  & \multicolumn{1}{c}{$\sigma = 0.25$} &  \multicolumn{1}{c}{$\sigma = 0.3$}                 
& \multicolumn{1}{c}{$\sigma = 0.2$}  & \multicolumn{1}{c}{$\sigma = 0.25$} &  \multicolumn{1}{c}{$\sigma = 0.3$} \\ 
\multirow{-3}{*}{Method} & 10.902 &	8.963 	&7.380 	&10.902 &	8.963 &	7.380& 	10.902 &	8.963& 	7.380 
 \\
\hline
Median1 & 15.165 & 13.660 & 12.395 & 14.276 & 13.106 & 12.084 & 13.307 & 12.562 & 11.850 \\
Median2 & 13.338 & 12.960 & 12.488 & 11.880 & 11.434 & 11.002 & 10.749 & 10.668 & 10.484 \\
ARMA3 & 8.646 & 8.008 & 7.336 & 11.140 & 9.543 & 8.156 & 10.902 & 8.963 & 7.380 \\
ARMA4 & 12.115 & 10.859 & 9.691 & 4.913 & 4.410 & 3.886 & 10.902 & 8.963 & 7.380 \\
ARMA5 & 11.936 & 10.994 & 10.058 & 12.540 & 11.001 & 9.659 & 10.902 & 8.963 & 7.380 \\
GCN & 11.667 & 11.022 & 10.522 & 10.931 & 9.551 & 9.489 & 10.987 & 9.867 & 9.366 \\
ChebNet & 19.265 & 17.739 & 16.576 & 19.997 & 18.951 & 18.136 & 18.243 & 17.827 & 16.962 \\
GAT & 13.953 & 12.473 & 11.517 & 13.884 & 11.594 & 10.761 & 11.812 & 11.794 & 11.093 \\
\multirow{2}{*}{G-adj1} & 14.449 & 13.056 & 11.944 & 13.946 & 12.367 & 11.117 & 14.249 & 12.816 & 11.675 \\
 & (0.93) & (0.93) & (0.93) & (1.04) & (1.04) & (1.04) & (0.98) & (0.98) & (0.98) \\
\multirow{2}{*}{G-adj2} & 14.461 & 13.065 & 11.953 & 13.961 & 12.376 & 11.124 & 14.256 & 12.822 & 11.682 \\
 & (0.928) & (0.931) & (0.933) & (1.044) & (1.043) & (1.043) & (0.979) & (0.981) & (0.983) \\
\multirow{2}{*}{G-lap1} & 14.042 & 12.593 & 11.475 & 14.056 & 12.660 & 11.570 & 14.910 & 13.584 & 12.515 \\
 & (0.99) & (0.99) & (0.99) & (1.00) & (1.00) & (0.99) & (0.89) & (0.89) & (0.89) \\
\multirow{2}{*}{G-lap2} & 14.047 & 12.597 & 11.480 & 14.066 & 12.667 & 11.576 & 14.916 & 13.590 & 12.522 \\
 & (0.989) & (0.992) & (0.995) & (1.006) & (0.999) & (0.989) & (0.888) & (0.889) & (0.889) \\
\multirow{2}{*}{G-row-norm-adj1} & 14.420 & 13.003 & 11.871 & 14.291 & 12.709 & 11.504 & 13.887 & 12.589 & 11.565 \\
 & (1.02) & (1.02) & (1.01) & (1.10) & (1.08) & (1.05) & (1.03) & (1.05) & (1.05) \\
\multirow{2}{*}{G-row-norm-adj2} & 14.424 & 13.010 & 11.877 & 14.302 & 12.716 & 11.510 & 14.018 & 12.669 & 11.614 \\
 & (1.019) & (1.015) & (1.012) & (1.097) & (1.080) & (1.055) & (1.105) & (1.093) & (1.081) \\
\multirow{2}{*}{G-sym-norm-adj1} & 14.423 & 12.988 & 11.839 & 14.535 & 13.058 & 11.880 & 14.301 & 12.903 & 11.805 \\
 & (1.02) & (1.02) & (1.02) & (1.07) & (1.07) & (1.07) & (1.08) & (1.07) & (1.07) \\
\multirow{2}{*}{G-sym-norm-adj2} & 14.432 & 12.994 & 11.844 & 14.550 & 13.073 & 11.894 & 14.307 & 12.907 & 11.809 \\
 & (1.017) & (1.019) & (1.020) & (1.068) & (1.067) & (1.066) & (1.081) & (1.074) & (1.065) \\
\bottomrule
\end{tabular}
\end{sidewaystable}

\begin{sidewaystable}[!htbp]

\footnotesize
\centering
\setlength{\tabcolsep}{5pt}
\renewcommand{\arraystretch}{1.4}
\ContinuedFloat
\caption{SNR results for transform order pair learning experiment for SST(Continued)}
\begin{tabular}{cccccccccc}
\toprule
& \multicolumn{3}{c}{2-NN} & \multicolumn{3}{c}{3-NN} & \multicolumn{3}{c}{5-NN} \\ \cline{2-10}
& \multicolumn{1}{c}{$\sigma = 0.2$}  & \multicolumn{1}{c}{$\sigma = 0.25$} &  \multicolumn{1}{c}{$\sigma = 0.3$}                 
& \multicolumn{1}{c}{$\sigma = 0.2$}  & \multicolumn{1}{c}{$\sigma = 0.25$} &  \multicolumn{1}{c}{$\sigma = 0.3$}                 
& \multicolumn{1}{c}{$\sigma = 0.2$}  & \multicolumn{1}{c}{$\sigma = 0.25$} &  \multicolumn{1}{c}{$\sigma = 0.3$} \\ 
\multirow{-3}{*}{Method} & 10.902 &	8.963 	&7.380 	&10.902 &	8.963 &	7.380& 	10.902 &8.963& 	7.380  \\
\hline
\multirow{2}{*}{G-norm-lap1} & 14.401 & 12.987 & 11.879 & 13.915 & 12.445 & 11.301 & 14.295 & 12.881 & 11.770 \\
 & (1.02) & (1.03) & (1.04) & (0.99) & (0.99) & (0.99) & (1.06) & (1.06) & (1.04) \\
\multirow{2}{*}{G-norm-lap2} & 14.406 & 12.993 & 11.884 & 13.927 & 12.453 & 11.308 & 14.302 & 12.887 & 11.775 \\
 & (1.018) & (1.027) & (1.039) & (0.991) & (0.992) & (0.993) & (1.064) & (1.055) & (1.045) \\
\multirow{2}{*}{J-adj1} & 20.562 & 19.144 & 17.960 & 20.876 & 19.555 & 18.452 & 20.790 & 19.434 & 18.315 \\
 & (0.90,$-1.00$) & (0.90,$-1.00$) & (0.90,$-1.00$) & (1.97,$-1.00$) & (1.97,$-1.00$) & (1.97,$-1.00$) & ($-1.01$,$-1.00$) & ($-1.01$,$-1.00$) & ($-1.01$,$-1.00$) \\
\multirow{2}{*}{J-adj2} & \textbf{20.694} & \textbf{19.271} & \textbf{18.081} & 21.066 & 19.727 & \textbf{18.610} & \textbf{20.970} & \textbf{19.597} & \textbf{18.464} \\
 & \textbf{(0.898,1.000)}& \textbf{(0.899,1.000)} & \textbf{(0.900,$-$1.000)} & (1.975,$-1.000$) & (1.972,$-1.000$) & \textbf{(1.968,$-$1.000)} & \textbf{($-$1.006,$-$1.000)} & \textbf{($-$1.007,$-$1.000)} & \textbf{($-$1.008,1.000)} \\
\multirow{2}{*}{J-lap1} & 20.202 & 18.692 & 17.455 & 20.899 & 19.513 & 18.369 & 20.814 & 19.459 & 18.336 \\
 & (0.98,$-1.00$) & (0.98,$-1.00$) & (0.99,$-1.00$) & ($-1.69$,$-1.00$) & ($-1.71$,$-1.00$) & ($-1.72$,$-1.00$) & (0.88,$-1.00$) & (0.88,1.00) & (0.88,$-1.00$) \\
\multirow{2}{*}{J-lap2} & 20.318 & 18.792 & 17.537 &\textbf{ 21.138} & \textbf{19.747} & 18.600 & 20.954 & 19.569 & 18.423 \\
 & (0.984,$-1.000$) & (0.985,$-1.000$) & (0.987,$-1.000$) & \textbf{($-$1.690,1.000)} & \textbf{($-$1.704,1.000)} & ($-$1.721,1.000) & (0.880,1.000) & (0.881,$-1.000$) & (0.882,$-1.000$) \\
\multirow{2}{*}{J-row-norm-adj1} & 20.420 & 19.015 & 17.846 & 20.789 & 19.264 & 17.979 & 19.892 & 18.354 & 17.171 \\
 & (0.92,$-1.00$) & (0.92,$-1.00$) & (0.92,$-1.00$) & (1.11,$-1.00$) & (1.11,$-1.00$) & (1.11,1.00) & (2.00,$-1.00$) & (1.01,$-1.00$) & (1.01,$-1.00$) \\
\multirow{2}{*}{J-row-norm-adj2} & 20.578 & 19.154 & 17.964 & 20.931 & 19.396 & 18.158 & 20.185 & 18.714 & 17.514 \\
 & (0.914,$-1.000$) & (0.914,1.000) & (0.916,$-1.000$) & (1.114,$-1.000$) & (1.115,$-1.000$) & (1.617,$-1.000$) & (1.117,1.000) & (1.121,$-1.000$) & (1.124,$-0.999$) \\
\multirow{2}{*}{J-sym-norm-adj1} & 20.421 & 18.972 & 17.782 & 20.445 & 18.980 & 17.773 & 20.253 & 18.786 & 17.580 \\
 & (1.01,$-1.00$) & (1.01,$-1.00$) & (1.01,1.00) & ($-1.92$,$-1.00$) & (1.07,$-1.00$) & (1.07,$-1.00$) & (1.09,$-1.00$) & (1.09,1.00) & (1.09,$-1.00$) \\
\multirow{2}{*}{J-sym-norm-adj2} & 20.545 & 19.074 & 17.866 & 20.695 & 19.127 & 17.898 & 20.355 & 18.875 & 17.661 \\
 & (1.007,$-1.000$) & (1.008,1.000) & (1.008,1.000) & ($-1.918$,1.000) & (1.072,$-1.000$) & (1.072,$-1.000$) & (1.089,$-1.000$) & (1.093,$-1.000$) & (1.096,$-1.000$) \\
\multirow{2}{*}{J-norm-lap1} & 20.442 & 18.989 & 17.797 & 20.448 & 19.052 & 17.896 & 20.624 & 19.218 & 18.033 \\
 & (1.01,$-1.00$) & (1.01,$-1.00$) & (1.01,$-1.00$) & (1.32,$-1.00$) & (1.31,$-1.00$) & (1.30,$-1.00$) & (1.99,$-1.00$) & (1.99,$-1.00$) & (1.98,$-1.00$) \\
\multirow{2}{*}{J-norm-lap2} & 20.606 & 19.091 & 17.882 & 20.658 & 19.260 & 18.108 & 20.811 & 19.403 & 18.212 \\
 & ($-0.117$,$-1.000$) & (1.011,1.000) & (1.013,1.000) & (1.317,$-1.000$) & (1.313,$-1.000$) & (1.308,$-1.000$) & (1.987,$-1.000$) & (1.981,$-1.000$) & (1.975,$-1.000$) \\
 \bottomrule
\end{tabular}
\end{sidewaystable}

\begin{sidewaystable}[!htbp]
\caption{SNR results for transform order pair learning experiment for PM-2.5}
\label{tab:PM2.5}
\footnotesize
\centering
\setlength{\tabcolsep}{18pt}
\renewcommand{\arraystretch}{1.4}
\begin{tabular}{cccccccccc}
\toprule
& \multicolumn{3}{c}{2-NN} & \multicolumn{3}{c}{3-NN} & \multicolumn{3}{c}{5-NN} \\ \cline{2-10}
& \multicolumn{1}{c}{$\sigma = 0.15$}  & \multicolumn{1}{c}{$\sigma = 0.2$} &  \multicolumn{1}{c}{$\sigma = 0.25$}                 
& \multicolumn{1}{c}{$\sigma = 0.15$}  & \multicolumn{1}{c}{$\sigma = 0.2$} &  \multicolumn{1}{c}{$\sigma = 0.25$}                 
& \multicolumn{1}{c}{$\sigma = 0.15$}  & \multicolumn{1}{c}{$\sigma = 0.2$} &  \multicolumn{1}{c}{$\sigma = 0.25$} \\ 
\multirow{-3}{*}{Method} & 4.463 &	1.964 &	0.026 &	4.463 &	1.964 &	0.026& 	4.463 &	1.964& 	0.026 
 \\
\hline
Median1 & 8.953 & 7.337 & 5.825 & 9.000 & 7.555 & 6.225 & 8.413 & 7.405 & 6.378 \\
Median2 & 8.802 & 7.955 & 6.974 & 8.900 & 8.272 & 7.508 & 7.933 & 7.649 & 7.160 \\
ARMA3 & 4.817 & 2.321 & 0.472 & 4.747 & 2.255 & 0.320 & 4.463 & 1.964 & 0.026 \\
ARMA4 & 5.115 & 2.633 & 0.701 & 4.985 & 2.546 & 0.636 & 4.463 & 1.964 & 0.026 \\
ARMA5 & 5.598 & 3.134 & 1.211 & 5.008 & 2.571 & 0.662 & 4.463 & 1.964 & 0.026 \\
GCN & 8.645 & 7.228 & 6.153 & 8.875 & 7.753 & 6.555 & 8.131 & 7.139 & 6.274 \\
ChebyNet & 8.853 & 7.683 & 6.796 & 8.800 & 7.046 & 6.161 & 8.086 & 6.984 & 6.375 \\
GAT & 8.994 & 7.551 & 6.515 & 8.215 & 7.891 & 6.848 & 7.959 & 7.455 & 6.639 \\
\multirow{2}{*}{G-adj1} & 8.141 & 6.639 & 5.564 & 9.370 & 8.066 & 7.063 & 9.966 & 8.690 & 7.671 \\
 & (1.04) & (1.04) & (1.05) & (0.93) & (0.92) & (0.92) & (0.97) & (0.96) & (0.96) \\
\multirow{2}{*}{G-adj2} & 8.149 & 6.647 & 5.571 & 9.384 & 8.083 & 7.082 & 9.985 & 8.745 & 7.703 \\
 & (1.039) & (1.043) & (1.051) & (0.925) & (0.921) & (0.918) & (0.968) & (0.963) & (0.959) \\
\multirow{2}{*}{G-lap1} & 9.175 & 7.655 & 6.511 & 9.827 & 8.410 & 7.337 & 9.776 & 8.390 & 7.332 \\
 & (1.01) & (1.02) & (1.02) & (1.02) & (1.02) & (1.02) & (1.00) & (1.00) & (1.00) \\
\multirow{2}{*}{G-lap2} & 9.195 & 7.674 & 6.527 & 9.839 & 8.422 & 7.350 & 9.791 & 8.412 & 7.348 \\
 & (1.010) & (1.017) & (1.025) & (1.017) & (1.018) & (1.018) & (1.001) & (1.001) & (0.999) \\
\multirow{2}{*}{G-row-norm-adj1} & 8.989 & 7.484 & 6.339 & 9.548 & 8.150 & 7.082 & 9.830 & 8.406 & 7.339 \\
 & (1.04) & (1.05) & (1.07) & (1.01) & (1.02) & (1.03) & (0.96) & (0.96) & (0.96) \\
\multirow{2}{*}{G-row-norm-adj2} & 9.110 & 7.573 & 6.402 & 9.559 & 8.162 & 7.093 & 9.871 & 8.441 & 7.371 \\
 & (1.074) & (1.082) & (1.088) & (1.013) & (1.023) & (1.029) & (0.962) & (0.963) & (0.963) \\
\multirow{2}{*}{G-sym-norm-adj1} & 9.046 & 7.548 & 6.412 & 9.929 & 8.553 & 7.476 & 10.272 & 8.898 & 7.818 \\
 & (1.06) & (1.07) & (1.07) & (0.97) & (0.97) & (0.97) & (0.97) & (0.97) & (0.97) \\
\multirow{2}{*}{G-sym-norm-adj2} & 9.058 & 7.559 & 6.422 & 9.941 & 8.568 & 7.493 & 10.301 & 8.933 & 7.854 \\
 & (1.064) & (1.069) & (1.072) & (0.970) & (0.970) & (0.970) & (0.972) & (0.973) & (0.974) \\
\bottomrule
\end{tabular}
\end{sidewaystable}
\begin{sidewaystable}[!htbp]

\ContinuedFloat
\caption{SNR results for transform order pair learning experiment for PM2.5(Continued)}
\footnotesize
\centering
\setlength{\tabcolsep}{5pt}
\renewcommand{\arraystretch}{1.4}
\begin{tabular}{cccccccccc}
\toprule
& \multicolumn{3}{c}{2-NN} & \multicolumn{3}{c}{3-NN} & \multicolumn{3}{c}{5-NN} \\ \cline{2-10}
& \multicolumn{1}{c}{$\sigma = 0.15$}  & \multicolumn{1}{c}{$\sigma = 0.2$} &  \multicolumn{1}{c}{$\sigma = 0.25$}                 
& \multicolumn{1}{c}{$\sigma = 0.15$}  & \multicolumn{1}{c}{$\sigma = 0.2$} &  \multicolumn{1}{c}{$\sigma = 0.25$}                 
& \multicolumn{1}{c}{$\sigma = 0.15$}  & \multicolumn{1}{c}{$\sigma = 0.2$} &  \multicolumn{1}{c}{$\sigma = 0.25$} \\ 
\multirow{-3}{*}{Method} & 4.463 &	1.964 &	0.026 &	4.463 &	1.964 &	0.026& 	4.463 &	1.964& 	0.026 
 \\
\hline
\multirow{2}{*}{G-norm-lap1} & 9.117 & 7.646 & 6.528 & 9.898 & 8.530 & 7.459 & 10.169 & 8.830 & 7.773 \\
 & (1.03) & (1.03) & (1.03) & (0.99) & (0.99) & (0.99) & (1.01) & (1.02) & (1.03) \\
\multirow{2}{*}{G-norm-lap2} & 9.130 & 7.658 & 6.541 & 9.912 & 8.546 & 7.475 & 10.199 & 8.862 & 7.804 \\
 & (1.033) & (1.030) & (1.025) & (0.994) & (0.992) & (0.990) & (1.015) & (1.023) & (1.028) \\
\multirow{2}{*}{J-adj1} & 10.835 & 9.678 & 8.809 & 11.220 & 10.127 & 9.336 & 11.215 & 10.257 & 9.592 \\
 & (1.05,$-$1.00) & (1.04,$-$1.00) & (1.04,$-$1.00) & (0.90,$-$1.00) & (0.90,$-$1.00) & (0.91,$-$1.00) & (0.98,$-$1.00) & (0.98,$-$1.00) & (0.98,$-$1.00) \\
\multirow{2}{*}{J-adj2} & 11.036 & 9.892 & 9.018 & 11.412 & 10.316 & 9.522 & 11.427 & 10.493 & 9.843 \\
 & (1.049,$-$1.000) & (1.042,$-$1.000) & (1.034,$-$1.000) & (0.898,$-$1.001) & (0.903,$-$1.000) & (0.908,$-$0.999) & (0.984,$-$0.999) & (0.981,$-$1.000) & (0.979,$-$1.000) \\
\multirow{2}{*}{J-lap1} & 11.252 & 10.234 & 9.469 & 11.445 & 10.439 & 9.717 & 11.379 & 10.405 & 9.723 \\
 & (1.01,$-$1.00) & (1.01,$-$1.00) & (1.02,$-$1.00) & (1.02,$-$1.00) & (1.01,$-$1.00) & (1.01,$-$1.00) & (1.00,$-$1.00) & (1.02,$-$1.00) & (1.02,$-$1.00) \\
\multirow{2}{*}{J-lap2} & \textbf{11.462} & \textbf{10.457} & \textbf{9.684} & 11.641 & 10.635 & 9.917 & 11.546 & 10.587 & 9.917 \\
 & \textbf{(1.005,$-$1.000)} & \textbf{(1.010,$-$1.000)} & \textbf{(1.016,$-$1.000)} & (1.018,1.001) & (1.011,$-$0.999) & (1.010,$-$0.999) & (0.993,$-$1.057) & (1.027,$-$1.000) & (1.021,$-$1.000) \\
\multirow{2}{*}{J-row-norm-adj1} & 11.028 & 10.003 & 9.249 & 11.229 & 10.279 & 9.632 & 11.158 & 10.161 & 9.446 \\
 & (1.00,$-$1.00) & (1.01,$-$1.00) & (1.01,$-$1.00) & (1.07,$-$1.00) & (1.08,$-$1.00) & (1.08,$-$1.00) & (0.98,$-$1.00) & (0.97,$-$1.00) & (0.97,$-$1.00) \\
\multirow{2}{*}{J-row-norm-adj2} & 11.374 & 10.389 & 9.638 & 11.441 & 10.490 & 9.809 & 11.461 & 10.424 & 9.694 \\
 & (1.099,$-$1.003) & (1.099,$-$1.002) & (1.097,$-$1.001) & (1.083,$-$1.000) & (1.086,$-$1.000) & (1.085,$-$1.000) & (0.959,$-$1.091) & (0.962,$-$1.003) & (0.964,$-$1.003) \\
\multirow{2}{*}{J-sym-norm-adj1} & 11.190 & 10.174 & 9.405 & 11.461 & 10.536 & 9.878 & 11.158 & 10.445 & 9.793 \\
 & (1.02,$-$1.00) & (1.02,$-$1.00) & (1.03,$-$1.00) & (0.95,$-$1.00) & (0.95,$-$1.00) & (0.94,$-$1.00) & (0.96,$-$1.00) & (0.96,$-$1.00) & (0.96,$-$1.00) \\
\multirow{2}{*}{J-sym-norm-adj2} & 11.426 & 10.426 & 9.651 & \textbf{11.670 }& \textbf{10.743} & \textbf{10.084} & \textbf{11.604} & \textbf{10.661 }& \textbf{10.028} \\
 & (1.021,$-$1.000) & (1.024,$-$1.000) & (1.029,$-$1.000) &\textbf{ (0.944,$-$1.006)} & \textbf{(0.943,$-$1.003)} & \textbf{(0.941,$-$1.004)} & \textbf{(0.961,$-$1.000)} &\textbf{ (0.961,$-$1.000)} & \textbf{(0.964,$-$1.000)} \\
\multirow{2}{*}{J-norm-lap1} & 11.173 & 10.172 & 9.415 & 11.352 & 10.423 & 9.758 & 11.295 & 10.329 & 9.676 \\
 & (1.02,$-$1.00) & (1.02,$-$1.00) & (1.02,$-$1.00) & (1.00,$-$1.00) & (1.00,$-$1.00) & (1.00,$-$1.00) & (1.02,$-$1.00) & (1.02,$-$1.00) & (1.02,$-$1.00) \\
\multirow{2}{*}{J-norm-lap2} & 11.425 & 10.429 & 9.663 & 11.537 & 10.619 & 9.962 & 11.518 & 10.535 & 9.904 \\
 & (1.017,$-$1.000) & (1.021,$-$1.000) & (1.024,$-$1.000) & (0.998,$-$1.061) & (0.999,$-$1.000) & (0.999,$-$1.000) & (1.017,$-$1.090) & (1.018,$-$1.000) & (1.021,$-$1.000) \\
\bottomrule
\end{tabular}
\end{sidewaystable}

\clearpage
\begin{sidewaystable}[!htbp]

\caption{SNR results for transform order pair learning experiment for COVID}
\label{tab:COVID}
\footnotesize
\centering
\setlength{\tabcolsep}{18pt}
\renewcommand{\arraystretch}{1.4}
\begin{tabular}{cccccccccc}
\toprule
& \multicolumn{3}{c}{2-NN} & \multicolumn{3}{c}{3-NN} & \multicolumn{3}{c}{5-NN} \\ \cline{2-10}
& \multicolumn{1}{c}{$\sigma = 0.1$}  & \multicolumn{1}{c}{$\sigma = 0.15$} &  \multicolumn{1}{c}{$\sigma = 0.2$}                 
& \multicolumn{1}{c}{$\sigma = 0.1$}  & \multicolumn{1}{c}{$\sigma = 0.15$} &  \multicolumn{1}{c}{$\sigma = 0.2$}                 
& \multicolumn{1}{c}{$\sigma = 0.1$}  & \multicolumn{1}{c}{$\sigma = 0.15$} &  \multicolumn{1}{c}{$\sigma = 0.2$} \\ 
\multirow{-3}{*}{Method} & 14.224 & 10.702 & 8.204 & 14.224 & 10.702 & 8.204 & 14.224 & 10.702 & 8.204 \\
\hline
Median1 & 12.390 & 10.949 &9.674 & 11.595&10.256 	&9.252 &	6.583 &	6.139 &	5.646 
\\
Median2 & 5.056 & 5.024 & 4.874 & 6.097 & 5.968 & 5.830 & 2.590 & 2.657 & 2.654 \\
ARMA3 & 11.536&	9.898&8.288&14.333 &11.110 &8.709&14.224& 10.702& 8.204 \\
ARMA4& 11.315&9.915&8.474&13.901&11.034 &8.776&14.224 &10.702 &8.204 \\
ARMA5& 10.634&9.666	&8.568&13.735 &11.053&	8.876&14.224 &10.702 &8.204  \\
GCN& 15.567 & 12.411 & 10.886 & 15.504 & 12.432 & 10.251 & 15.300 & 12.650 & 10.967 \\
ChebyNet & 19.234 & 18.124 & 16.306 & 19.372 & 17.504 & 16.072 & 20.733 & 18.122 & 15.505 \\
GAT & 15.626 & 12.605 & 10.765 & 15.645 & 12.627 & 10.785 & 15.659 & 12.641 & 10.774 \\
\multirow{2}{*}{G-adj1} & 16.886 & 14.051 & 12.136 & 16.437 & 13.567 & 11.671 & 16.655 & 13.912 & 12.063 \\
 & (1.17)&(1.17)&(1.17)&(1.04)&(1.06)&(1.07)&($-$1.18)&($-$1.18)&($-$1.18)\\
\multirow{2}{*}{G-adj2} & 16.916&14.077 &12.158& 16.509& 13.620& 11.711 &16.683&13.939&	12.089\\
 &(1.168)&(1.166)&(1.166)&(1.044)&(1.060)& (1.075)&($-$1.175)&($-$1.175)&($-$1.176) \\
\multirow{2}{*}{G-lap1} & 16.239 & 13.273 & 11.287 & 15.814 & 12.757 & 10.725 & 16.026 & 13.018 & 10.979 \\
 &(1.21) &(1.21) &(1.21) &($-$0.03)&($-$0.03) &	(0.94) &(1.22)&	(1.22)&	(1.22)\\
\multirow{2}{*}{G-lap2} & 16.321 & 13.347 & 11.354 & 15.862 & 12.801 & 10.790 & 16.099 & 13.084 & 11.039 \\
 & (1.207) &	(1.211) &(1.213)&($-$0.031) &($-$0.035)&(0.942)&(1.224)&(1.222)&(1.219)\\
\multirow{2}{*}{G-row-norm-adj1} & 16.076 & 13.237 & 11.297 & 15.752 & 12.800 & 10.904 & 15.975 & 12.983 & 11.047 \\
 & ($-$1.25) & (0.10) & (0.10) &( $-$0.01) &( $-$1.00) & ($-$1.00) & ($-$0.03) & ($-$0.91 )& ($-$0.91) \\
\multirow{2}{*}{G-row-norm-adj2} & 16.595 & 13.730 & 11.750 & 15.884 & 12.924 & 10.985 & 16.032 & 13.026 & 11.089 \\
 & (2.047)&(2.052) &( 2.057) &( $-$0.021) &( $-$1.035) &( $-$1.030) &( $-$0.035) &( $-$0.910) &( $-$0.909)\\
\multirow{2}{*}{G-sym-norm-adj1} & 15.911 & 13.073 & 11.185 & 16.600 & 13.718 & 11.713 & 17.021 & 14.185 & 12.291 \\
 & ($-$1.13 )&( $-$1.14) &( $-$1.15 )&( 1.34) &( 1.31) & (1.30) &( $-$1.82) &( $-$1.83) & ($-$1.84) \\
\multirow{2}{*}{G-sym-norm-adj2} & 16.109 & 13.213 & 11.262 & 16.684 & 14.180 & 12.300 & 17.116 & 14.259 & 12.347 \\
 & (2.086) &( 2.083) & ($-$1.151 )& (1.337) & ($-$2.344) & ($-$2.337) & ($-$1.824) & ($-$1.830 )&( $-$1.842) \\
\bottomrule
\end{tabular}
\end{sidewaystable}

\begin{sidewaystable}[!htbp]
\ContinuedFloat
\caption{SNR results for transform order pair learning experiment for COVID(Continued)}
\footnotesize
\centering
\setlength{\tabcolsep}{4pt}
\renewcommand{\arraystretch}{1.4}
\begin{tabular}{cccccccccc}
\toprule
& \multicolumn{3}{c}{2-NN} & \multicolumn{3}{c}{3-NN} & \multicolumn{3}{c}{5-NN} \\ \cline{2-10}
& \multicolumn{1}{c}{$\sigma = 0.1$}  & \multicolumn{1}{c}{$\sigma = 0.15$} &  \multicolumn{1}{c}{$\sigma = 0.2$}                 
& \multicolumn{1}{c}{$\sigma = 0.1$}  & \multicolumn{1}{c}{$\sigma = 0.15$} &  \multicolumn{1}{c}{$\sigma = 0.2$}                 
& \multicolumn{1}{c}{$\sigma = 0.1$}  & \multicolumn{1}{c}{$\sigma = 0.15$} &  \multicolumn{1}{c}{$\sigma = 0.2$} \\ 
\multirow{-3}{*}{Method} & 14.224 & 10.702 & 8.204 & 14.224 & 10.702 & 8.204 & 14.224 & 10.702 & 8.204 \\
\hline
\multirow{2}{*}{G-norm-lap1} & 16.171 &	13.299 &11.370 &16.184& 13.138 &11.079 &16.104& 13.272 &11.413\\
 &(0.73) &($-$1.02 )&($-$1.02) &(1.21) &(1.21) &(1.21 )&($-$1.70) &($-$1.73)&($-$1.77)\\
\multirow{2}{*}{G-norm-lap2} & 16.260 & 13.355 & 11.423 & 16.245 & 13.182 & 11.113 & 16.145 & 13.310 & 11.449 \\
 & (0.733) & ($-$1.017 )&( $-$1.016) &( 1.215) &( 1.215) &( 1.214 )&( $-$1.700) & ($-$1.731 )& ($-$1.768) \\
\multirow{2}{*}{J-adj1} & 24.374 & 21.547 & 19.575 & 23.795 & 20.694 & 18.548 & 24.134 & 21.039 & 18.907 \\
 & (1.18,$-$1.00) &( 1.17,$-$1.00) &( 1.17,$-$1.00) & (1.04,$-$1.00) &( 1.04,$-$1.00) &( 1.04,$-$1.00) &( $-$1.27,1.00 )& ($-$1.27,$-$1.00 )& ($-$1.27,$-$1.00) \\
\multirow{2}{*}{J-adj2} & 24.700 & \textbf{21.834} & \textbf{19.856} & \textbf{24.426} & \textbf{21.318} & \textbf{19.146} & 24.449 & 21.214 & 19.037 \\
 & (1.174, 1.000) & \textbf{(1.169, 1.000)} & \textbf{(1.164, 1.000)} & \textbf{(1.038, 1.000)} & \textbf{(1.040, 1.000)} & \textbf{(1.043, 1.000)} & ($-$1.275, 1.000 )& ($-$1.273, 1.000) &( $-$1.266, 1.000) \\
\multirow{2}{*}{J-lap1} & 23.953 &	20.711 &18.468 &23.457& 20.256& 18.032& 24.015& 20.704& 18.369 \\
 & (1.20,$-$1.00) & (1.20,$-$1.00) & (1.20,$-$1.00) &( $-$0.03,$-$1.00 )&($-$0.03,$-$1.00) &( $-$0.03,$-$1.00) &( 0.07,$-$1.00) &( 0.07,$-$1.00) & (0.08,$-$1.00) \\
\multirow{2}{*}{J-lap2} & 24.584 & 21.323 & 19.059 & 23.896 & 20.629 & 18.394 & 24.405 & 21.063 & 18.716 \\
 & (1.199, 1.000) & (1.199, $-$1.000 )& (1.200,1.000) & ($-$1.907, $-$1.000) &( $-$0.034,1.000) & ($-$0.035, $-$1.000) & ($-$2.008, 1.000) &( 0.072,$-$1.000 )&( 0.074,$-$1.000) \\
\multirow{2}{*}{J-row-norm-adj1} & 23.461 & 20.160 & 18.001 & 23.144 & 20.021 & 17.884 & 23.880 & 20.592 & 18.275 \\
 & ($-$1.69,$-$1.00) & ($-$0.08,$-$1.00) &( 0.11,$-$1.00) & (0.00,$-$1.00) & ($-$0.01,$-$1.00) & ($-$0.01,$-$1.00) &( $-$0.03,$-$1.00) &( $-$0.03,$-$1.00 )&( $-$0.03,$-$1.00) \\
\multirow{2}{*}{J-row-norm-adj2} & 24.004 & 20.649 & 18.339 & 23.841 & 20.640 & 18.406 & 24.303 & 21.019 & 18.684 \\
 & ($-$1.694, $-$1.000) &( $-$1.694, 1.000) &( 0.106,$-$1.000) &( $-$1.023,$-$1.000) &( $-$0.024, $-$1.000) & ($-$0.023, 1.000 )&( $-$0.032, $-$1.000) &( $-$0.036, 1.000) & ($-$0.037, 1.000 )\\
\multirow{2}{*}{J-sym-norm-adj1} & 23.408 & 20.230 & 18.109 & 23.821 & 20.594 & 18.286 & 24.116 & 20.902 & 18.730 \\
 & (0.09,$-$1.00) &( 0.08,$-$1.00 )& (0.07,$-$1.00 )& (1.35,$-$1.00) & (1.35,$-$1.00) & (1.36,$-$1.00) & (0.09,$-$1.00) & (1.91,$-$1.00 )& ($-$1.81,$-$1.00) \\
\multirow{2}{*}{J-sym-norm-adj2} & 23.908 & 20.780 & 18.694 & 24.260 & 21.040 & 18.888 &\textbf{ 24.531} & \textbf{21.368} & \textbf{19.250} \\
 & (0.088, $-$1.000) &( 2.088, $-$1.000 )& (2.088,$-$1.000) &( 1.079,1.000 )&( 1.080, $-$1.000 )& (1.356, $-$1.000) & \textbf{(0.088, $-$1.000)} &\textbf{ (1.906, 1.000 )}& \textbf{($-$1.813, $-$1.000 )}\\
\multirow{2}{*}{J-norm-lap1} & 24.290 & 21.034 & 18.723 & 23.901 & 20.563 & 18.225 & 23.821 & 20.893 & 18.866 \\
 & ($-$1.82,$-$1.00) & ($-$1.82,$-$1.00) &( $-$1.82,1.00) &( 0.90,$-$1.00) &( 0.90,$-$1.00) &( 0.90,1.00 )& ($-$1.67,$-$1.00) & ($-$1.68,1.00 )& ($-$1.68,1.00 )\\
\multirow{2}{*}{J-norm-lap2} & \textbf{24.714} & 21.431 & 19.115 & 24.141 & 21.277 & 18.922 & 24.233 & 21.273 & 19.219 \\
 & \textbf{($-$1.825, 1.000)} & ($-$1.824, 1.000) &( $-$1.824, $-$1.000) &( 1.765, $-$1.000) & (0.896, $-$1.000 )&( 0.896, $-$1.000) &( $-$1.677,1.000) & ($-$1.681, 1.000 )&( $-$1.683, 1.000) \\
\bottomrule
\end{tabular}
\end{sidewaystable} 

\begin{sidewaystable}[!htbp]
\centering
\footnotesize
\setlength{\tabcolsep}{10pt}
\renewcommand{\arraystretch}{2}
\caption{Comparison for Different Methods}
\label{tab:Comparsion of different methods}
\begin{tabular}{lcccccc}
\toprule
\textbf{Method} & \textbf{Parameter Count} & \textbf{Feature Type} & \textbf{Complexity} & \textbf{Prior Knowledge} & \textbf{Eigen Decomposition} \\
\midrule
ARMA & $P + Q + 1$ & Spatial feature & $O((P + Q )^2 + (PT + Q)E)$ & yes & no \\
Median & 0 & Spatio-temporal feature & $O(NTM\log M)$ & no & no \\
GCN & $\sum\limits_{l=1}^L (F_l {F'_l} + F'_l)$ & Spatial feature & $O(NFF' + EF)$ & yes & no \\
ChebyNet & $\sum\limits_{l=1}^L ((K + 1) F_l {F'_l} + F'_l)$ & Spatial feature & $O(NFF' + KEF)$ & yes & no \\
GAT & $\sum\limits_{l=1}^L (H_l (F_l {F'_l} + 2F'_l) + H_l F'_l)$ & Spatial feature & $O(HNFF' + HEF)$ & yes & no \\
GFRFT-search & $\frac{b - a}{\Delta p} + 1$ & Spatial feature & $O(N^4)$ & yes & yes \\
GFRFT-learn & $(N + 2)n_{\text{exp}}$ & Spatial feature & $O(N^3)$ & yes & yes \\
JFRFT-search & $\left(\frac{b - a}{\Delta p} + 1\right) \left(\frac{c - d}{\Delta q} + 1\right)$ & Spatio-temporal feature & $O(N^3 + N^4 T^4)$ & yes & yes \\
JFRFT-learn & $(NT + 2)n_{\text{exp}}$ & Spatio-temporal feature & $O(N^3 + N^2 T^2)$ & yes & yes \\
\bottomrule
\end{tabular}

\vspace{0.5em}
\begin{minipage}{0.95\linewidth}
\footnotesize
\textbf{Note:} $L$: number of layers; $F$, $F'$: input and output dimensions of the $l$-th layer; $K$: number of Chebyshev polynomial orders; $H$: number of attention heads; $N$: number of nodes; $T$: temporal interval length; $M$: number of neighbors per node; $E$: number of edges; $\Delta p$, $\Delta q$: step sizes for GFRFT and DFRFT; $n_{\text{exp}}$: number of experimental runs under different initializations; $b$, $a$: maximum and minimum transform orders for GFRFT; $c$, $d$: maximum and minimum transform orders for DFRFT.
\end{minipage}

\end{sidewaystable}

\clearpage

Here, we consider using our method for the image enhancement task. In the REDS dataset, three real-world scenes were selected for the experiments. For each scene, three consecutive image frames were sampled, and each frame was uniformly resized to a resolution of 340×200. Each image was divided into non-overlapping patches of size 20×20. For each group of patches at the same spatial location across the three frames, a time-varying graph signal model across three time points was constructed.

Specifically, graph signals were formed based on the grayscale values of pixels within each patch. The graph structure was established using a 4-nearest neighbor strategy, from which the adjacency matrix was derived and used as the graph shift operator. The GFT matrix was then obtained through eigen decomposition of the shift operator. Subsequently, JFRFT-based adaptive filtering was applied to each time-varying graph signal. By minimizing the MSE loss function, the transform order pair and filter coefficients were iteratively optimized to recover corresponding patches across the three frames. We employ the Adam optimizer to train the model parameters, with an initial learning rate set to \(2 \times 10^{-3}\). A StepLR learning rate decay strategy is applied, reducing the learning rate by a factor of \(\gamma = 0.8\) every 500 iterations. The initial values of both the filter coefficients and the transformation order pair are set to 1. This process was performed independently for all patches, and the recovered patches were then reassembled to reconstruct the complete image.

\begin{figure}[htbp]
\footnotesize
\centering
\adjustbox{max width=\textwidth}{
\begin{tabular}{ccc}
\includegraphics[height=2.8cm]{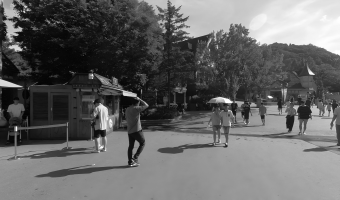} & 
\includegraphics[height=2.8cm]{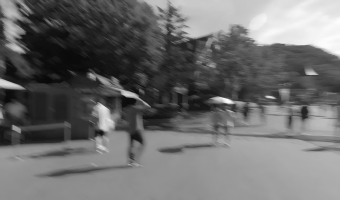} & 
\includegraphics[height=2.8cm]{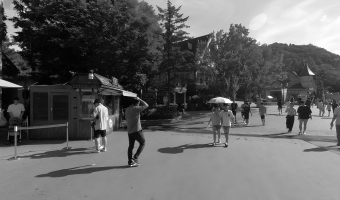} \\
\includegraphics[height=2.8cm]{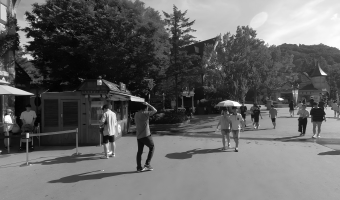} & 
\includegraphics[height=2.8cm]{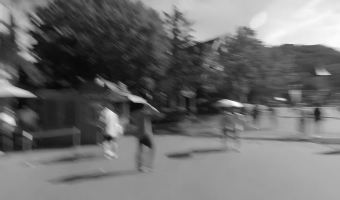} & 
\includegraphics[height=2.8cm]{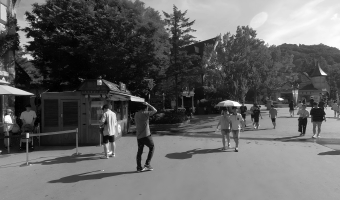} \\
\includegraphics[height=2.8cm]{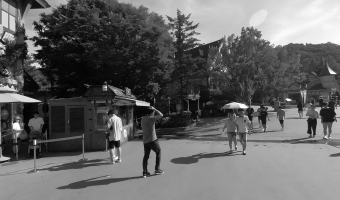} & 
\includegraphics[height=2.8cm]{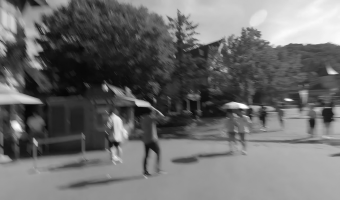} & 
\includegraphics[height=2.8cm]{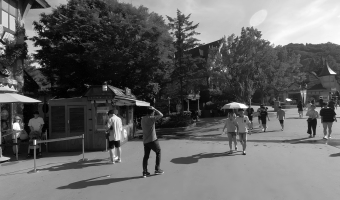} \\
Original &Blurred &Recovered \\
\end{tabular}
}
\caption{Recovering results on the REDSA dataset}
\label{fig:redsa}
\end{figure}

\begin{figure}[H]
\footnotesize
\centering
\adjustbox{max width=\textwidth}{
\begin{tabular}{ccc}
\includegraphics[height=2.8cm]{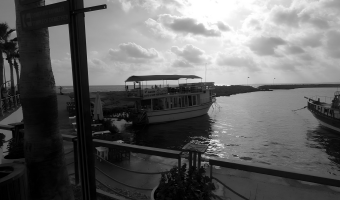} & 
\includegraphics[height=2.8cm]{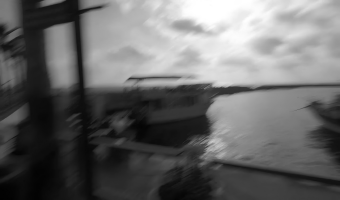} & 
\includegraphics[height=2.8cm]{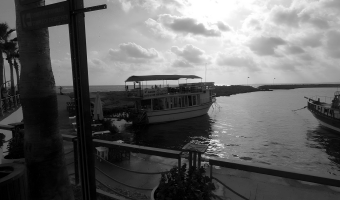} \\
\includegraphics[height=2.8cm]{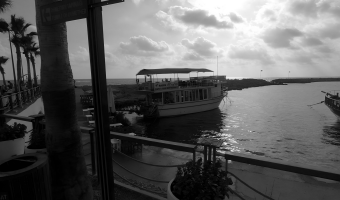} & 
\includegraphics[height=2.8cm]{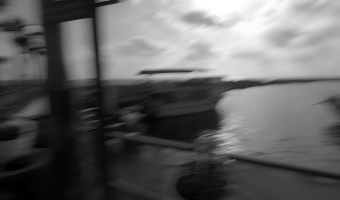} & 
\includegraphics[height=2.8cm]{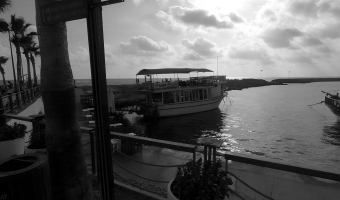} \\
\includegraphics[height=2.8cm]{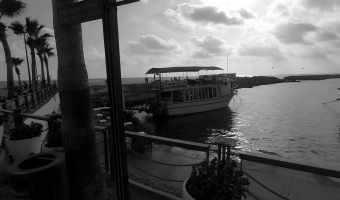} & 
\includegraphics[height=2.8cm]{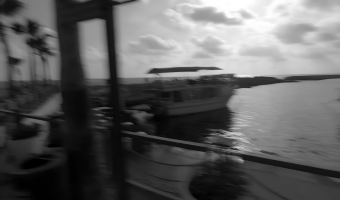} & 
\includegraphics[height=2.8cm]{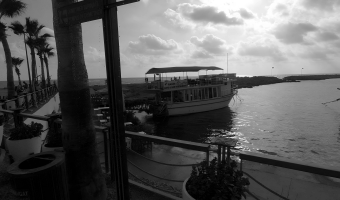} \\
Original & Blurred &Recovered \\
\end{tabular}
}
\caption{Recovering results on the REDSB dataset}
\label{fig:redsB}
\end{figure}

\begin{figure}[htbp]
\footnotesize
\centering
\adjustbox{max width=\textwidth}{
\begin{tabular}{ccc}
\includegraphics[height=2.8cm]{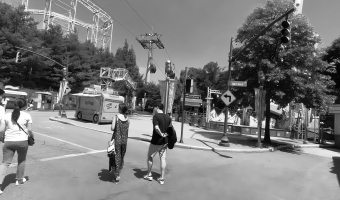} & 
\includegraphics[height=2.8cm]{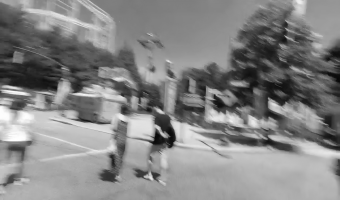} & 
\includegraphics[height=2.8cm]{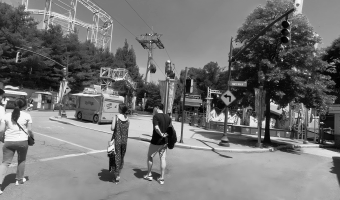} \\
\includegraphics[height=2.8cm]{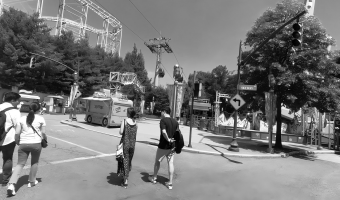} & 
\includegraphics[height=2.8cm]{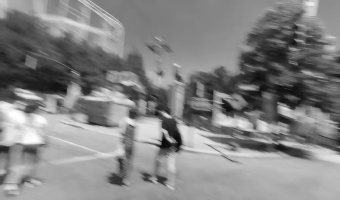} & 
\includegraphics[height=2.8cm]{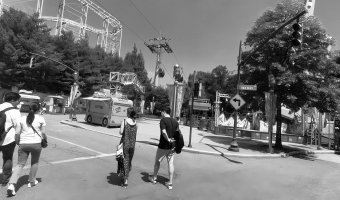} \\
\includegraphics[height=2.8cm]{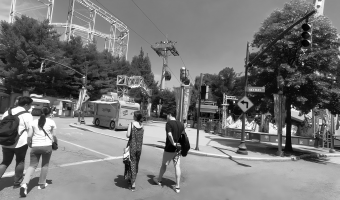} & 
\includegraphics[height=2.8cm]{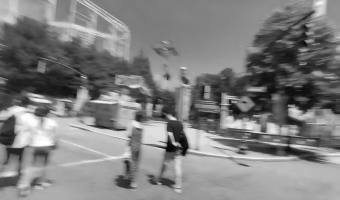} & 
\includegraphics[height=2.8cm]{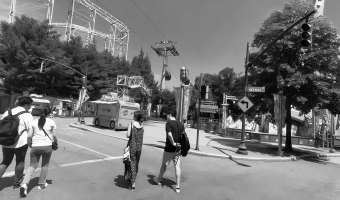} \\
Original & Blurred & Recovered \\
\end{tabular}
}
\caption{Recovering results on the REDSC dataset}
\label{fig:redsC}
\end{figure}

Figure~\ref{fig:redsa}, Figure~\ref{fig:redsB} and Figure~\ref{fig:redsC} present a visual comparison between the blurred and recovered images for the three selected scenes. Table~\ref{REDS} reports the quantitative evaluation results in terms of MSE, PSNR and SSIM, demonstrating the effectiveness of the proposed method in improving image quality.

\begin{table}[htbp]
\footnotesize
\centering
\setlength{\tabcolsep}{22pt}
\renewcommand{\arraystretch}{1.05}
\caption{Quantitative Recovering Results (MSE/PSNR/SSIM) on REDS Dataset}
\begin{tabular}{ccccc}
\toprule
\textbf{Dataset} & \textbf{Frame} & \textbf{MSE} & \textbf{PSNR} & \textbf{SSIM} \\
\midrule
\multirow{3}{*}{REDSA} & Frame1 & 4.517 & 41.582 & 0.9947 \\
                       & Frame2 & 5.307 & 40.882 & 0.9940 \\
                       & Frame3 & 3.966 & 42.148 & 0.9962 \\
\midrule
\multirow{3}{*}{REDSB} & Frame1 & 0.799 & 49.105 & 0.9967 \\
                       & Frame2 & 1.194 & 47.360 & 0.9955 \\
                       & Frame3 & 1.153 & 47.513 & 0.9953 \\
\midrule
\multirow{3}{*}{REDSC} & Frame1 & 5.297 & 40.890 & 0.9990 \\
                       & Frame2 & 1.196 & 47.355 & 0.9992 \\
                       & Frame3 & 1.573 & 46.163 & 0.9993 \\
\bottomrule
\label{REDS}
\end{tabular}
\end{table}

\clearpage

\subsubsection{Computational Cost}
In the grid search method based on the JFRFT, the graph shift operator and its corresponding GFT matrix must first undergo Jordan decomposition, which has a computational complexity of $\mathcal{O}(N^3)$, where $N$ denotes the number of nodes in the graph. For each pair of transform orders, the complexity of computing the corresponding filter coefficients is $\mathcal{O}(N^4T^4)$, where $T$ represents the temporal interval length. Thus, the overall computational complexity of this method is $\mathcal{O}(N^3 + N^4T^4)$. In contrast, the method proposed in this paper also requires Jordan decomposition of the graph shift operator and GFT matrix. However, during training, it dynamically updates parameters in each iteration, resulting in a per-epoch computational complexity of only $\mathcal{O}(N^2T^2)$. Consequently, the total complexity is reduced to $\mathcal{O}(N^3 + N^2T^2)$, demonstrating superior computational efficiency compared to the conventional grid search method.

To validate the above analysis, Table~\ref{fuzadu} reports the runtime required to complete the full denoising process for both methods on SST dataset of varying scales. For the grid search method, the transform order pair was searched within the range $[-2, 2]$ with a step size of $0.1$. As shown in the table, the proposed method consistently exhibits higher computational efficiency across all experimental settings. All experiments were conducted using MATLAB R2024a, Python 3.9.21, and PyTorch 2.6.0. The experiments were run on a 12th Gen Intel(R) Core(TM) i5-12600KF processor (3.70 GHz). The code is available at https://github.com/yezhiqiu258/LJFRFT.git.

\begin{table}[htbp]
\footnotesize
\centering
\setlength{\tabcolsep}{20pt}
\renewcommand{\arraystretch}{1.3}
\caption{Runtime comparison of JFRFT-search and JFRFT-learn under different scales}
\begin{tabular}{lccc}
\toprule
\textbf{Method} & $N{=}10, T{=}10$ & $N{=}15, T{=}15$ & $N{=}20, T{=}20$ \\
\midrule
\textbf{JFRFT-search} & 1280s & $4.32 \times 10^4$s & $1.40 \times 10^6$s \\
\textbf{JFRFT-learn}  & 990s  & 1200s               & 1950s \\
\bottomrule
\label{fuzadu}
\end{tabular}
\end{table}

\section{Conclusion}
In summary, this study developed a hyper-differential form of JFRFT, expanding its theoretical framework and enhancing its applicability to time-varying graph signal processing. To address the limitations of existing GFRFT-based and JFRFT-based Wiener filtering methods, the proposed approach leverages the newly defined differentiability with respect to transform order pair, embeds transform order pair and filter coefficients as learnable parameters within a neural network and achieves adaptive learning through gradient backpropagation. This method not only accounts for the temporal characteristics of time-varying graph signals but also overcomes the constraints of conventional grid search, considerably improving denoising efficiency. Moreover, the model can directly learn the optimal transform order pair from data. 

Experiments conducted on multiple synthetic and real-world datasets demonstrate that the proposed method substantially enhances the SNR in denoising tasks for time-varying graph signals. These findings underscore the effectiveness and robustness of the filtering approach, offering a more efficient solution. By integrating JFRFT with machine learning, this method holds potential for broader applications, including graph neural networks and time series prediction.

Although the proposed method demonstrates strong performance in terms of computational efficiency and denoising effectiveness, it still has the following limitations:

{\begin{enumerate}
    \item 
        \textbf{Dependence on Jordan Decomposition:}  
        The method relies on an exact Jordan decomposition of the graph shift operator. For large-scale graphs or when the shift operator is nearly defective, numerical instability during decomposition may adversely affect the filtering performance and reduce the robustness of the model.

    \item
        \textbf{Sensitivity to Hyperparameters:}  
        The proposed approach adopts an iterative optimization strategy for parameter updating. While this significantly reduces computational complexity compared to grid search, the final performance may still be influenced by the choice of initial values, learning rates, and other hyperparameters.

    \item
        \textbf{Limited Application Scope:}  
        The current framework is primarily validated on denoising tasks for time-varying graph signals and assumes prior knowledge of the graph signal characteristics. Its generalization to other graph signal processing tasks, such as classification, prediction, reconstruction, or compression, remains to be explored in future work.
    
\end{enumerate}}

\bibliographystyle{elsarticle-num}
\bibliography{reference}

\end{sloppypar}
\end{document}